\documentclass[times,twocolumn,final]{elsarticle}

\usepackage{medima}
\usepackage{framed,multirow}

\usepackage{amssymb}
\usepackage{latexsym}

\usepackage{url}
\usepackage{xcolor}

\usepackage{hyperref}

\usepackage{algorithm}
\usepackage[noend]{algpseudocode} 
\usepackage{bm}             
\usepackage{booktabs}       
\usepackage{amsfonts}       
\usepackage{nicefrac}       
\usepackage{microtype}      
\usepackage{ulem}           

\definecolor{newcolor}{rgb}{.8,.349,.1}

\journal{Medical Image Analysis}

\begin{document}

\verso{LaLonde \textit{et~al.}}

\begin{frontmatter}

\title{Capsules for Biomedical Image Segmentation}%

\author[1]{Rodney LaLonde}
\author[2]{Ziyue Xu}
\author[3]{Ismail Irmakci}
\author[4]{Sanjay Jain}
\author[1]{Ulas Bagci\corref{cor1}}
\cortext[cor1]{Corresponding author: 
  Tel.: +1-407-823-1047;
  fax: +1-407-823-0594;}
\ead{bagci@crcv.ucf.edu}

\address[1]{Center for Research in Computer Vision (CRCV), University of Central Florida (UCF), Orlando, FL}
\address[2]{Nvidia, Bethesda, MD}
\address[3]{Ege University, Izmir, Turkey}
\address[4]{Johns Hopkins University, Baltimore, MD}

\received{5 March 2020}
\finalform{25 August 2020}
\accepted{23 October 2020}
\availableonline{5 November 2020}

\newcommand{\RL}[1]{{\color{black} #1}}
\newcommand{\ADD}[1]{{\color{black} #1}}
\DeclareRobustCommand{\DEL}[1]{\texorpdfstring{{\color{black} \sout{#1}}}{#1}}

\begin{abstract}
Our work expands the use of capsule networks to the task of object segmentation for the first time in the literature. This is made possible via the introduction of \textit{locally-constrained routing} and \textit{transformation matrix sharing}, which reduces the parameter/memory burden and allows for the segmentation of objects at large resolutions. To compensate for the loss of global information in constraining the routing, we propose the concept of \textit{``deconvolutional'' capsules} to create a deep encoder-decoder style network, called \textbf{\textit{SegCaps}}. We extend the masked reconstruction regularization to the task of segmentation and perform thorough ablation experiments on each component of our method. The proposed convolutional-deconvolutional capsule network, \textit{SegCaps}, shows state-of-the-art results while using a fraction of the parameters of popular segmentation networks. To validate our proposed method, we perform experiments segmenting pathological lungs from clinical and pre-clinical thoracic computed tomography (CT) scans and segmenting muscle and adipose (fat) tissue from magnetic resonance imaging (MRI) scans of human subjects' thighs. Notably, our experiments in lung segmentation represent the largest-scale study in pathological lung segmentation in the literature, where we conduct experiments across five extremely challenging datasets, containing both clinical and pre-clinical subjects, and nearly $2000$ computed-tomography scans. Our newly developed segmentation platform outperforms other methods across all datasets while utilizing  less than $5$\% of the parameters in the popular \textit{U-Net} for biomedical image segmentation.Further, we demonstrate capsules' ability to generalize to unseen  rotations/reflections on natural images.
\end{abstract}


\end{frontmatter}


\newcommand{\RL}[1]{{\color{red} #1}}
\newcommand{\ADD}[1]{{\color{black} #1}}
\DeclareRobustCommand{\DEL}[1]{\texorpdfstring{{\color{black} \sout{#1}}}{#1}}

\section{Introduction}\label{sec:intro}
\ADD{The task of segmenting objects from images can be formulated as a joint object recognition and delineation problem. The goal in \textit{recognition} is to locate an object's presence in an image, whereas delineation attempts to draw the object's spatial extent and composition \citep{bagci2012hierarchical}. Solving these tasks jointly (or sequentially) results in partitions of non-overlapping, connected regions, homogeneous with respect to some signal characteristics. Object segmentation is an inherently difficult task; apart from recognizing the object, we also have to label that object at the pixel level, which is an ill-posed problem. 

Segmentation is of significant importance in biomedical image analysis, aiding systems focused on localizing pathologies~\mbox{\citep{Elbaz}}, tracking disease progression~\mbox{\citep{xucurrent}}, characterizing anatomical structure and defects~\mbox{\citep{Farag}}, and many more~\mbox{\citep{Elnakib}}. Due to its significance is many applications, segmentation is an essential part of most computer-aided diagnosis (CAD) systems, where the functionality of such systems can be heavily dependent on the accuracy of the segmentation module. Medical image segmentation brings its own set of unique challenges. Many anatomical structures vary significantly across individuals, with the presence of pathologies adding an additional layer of variation and complexity. Further, scanner artifacts and other noise can make the segmentation suboptimal. Recently convolutional neural network (CNN) methodologies have dominated the segmentation field, both in computer vision and medical image segmentation, most notably \textit{U-Net for biomedical image segmentation}~\citep{ronneberger2015u}, due to their remarkable predictive performance.}

\subsection{Drawbacks of CNNs and How Capsules Solve Them} \label{sec:cnn_shortcomings}

The CNNs, despite showing remarkable flexibility and performance in a wide range of computer vision tasks, do come with their own set of flaws. Due to the scalar and additive nature of neurons in CNNs, neurons at any given layer of a network are ambivalent to the spatial relationships of neurons within their kernel of the previous layer, and thus within their effective receptive field of the given input. \ADD{Feature maps in CNNs only contain scalar values, whether or not a given feature is present at each scalar location. These maps are created from one layer to the next by multiplying each previous layer feature map by a set of kernel, then summing their activations to designate the presence/absence of the next higher-level feature. Since CNNs only have the ability to add presence activations within local kernels, higher-level neurons can only identify features within their effective receptive fields, but they cannot describe those feature in any way (\textit{e.g.} precise location information, pose, deformation, etc.)} To address this significant shortcoming, \citet{sabour2017dynamic} introduced the idea of \textit{capsule networks}, where information at the neuron level is stored as vectors, rather than scalars. These vectors contain information about:
\begin{enumerate}
\item spatial orientation and location information,
\item magnitude/prevalence, and
\item other attributes of the extracted feature
\end{enumerate}
represented by each capsule type of that layer. \ADD{This solves the previous issue of precise spatial localization in CNNs because capsule vectors can now additionally rate-code the exact position within the effective receptive field of the already place-coded vectors within the dimensions of those vectors.} These sets of neurons, henceforth referred to as capsule types, are then ``routed'' to capsules in the next layer via a \textit{dynamic routing algorithm} which takes into account the agreement between these capsule vectors, thus forming meaningful part-to-whole relationships not found in standard CNNs.

\textbf{The overall goal} of this study is to extend capsule networks and the dynamic routing algorithm to accomplish the task of object segmentation for the first time in the literature. We hypothesize that capsules can be used effectively for object segmentation with high accuracy and heightened efficiency compared to the state-of-the-art segmentation methods. To show the efficacy of the capsules for object segmentation, we choose a challenging application of pathological lung segmentation from computed tomography (CT) scans, where we have analyzed the largest-scale study of data obtained from both clinical and pre-clinical subjects, comprising nearly $2000$ CT scans across five datasets\ADD{, and muscle and adipose (fat) tissue segmentation from magnetic resonance imaging (MRI) scans of three different contrasts obtained from a cohort of $50$ patients ($150$ scans)}.  We chose pathological lung segmentation for its obvious life-saving potential and unique challenges such as high intra-class variation, noise, artifacts and abnormalities, and other reasons discussed in Section~\ref{sec:related}. \ADD{The additional experiments on muscle and adipose (fat) tissue segmentation compliment these first experiments both in the modality of the imaging technology used (MRI vs. CT) and anatomical structure.} 
To further demonstrate the general applicability of our methods, we also provide proof-of-concept results for rotations/reflections on standard computer vision images showing\ADD{, the ability of a capsule-based segmentation network to generalize to unseen poses of objects,} a strong motivation for choosing capsule networks over CNNs in segmentation applications.

\subsection{\ADD{Building Blocks of Capsules for Segmentation}} \label{sec:intro_caps}

\ADD{Performing object segmentation with a capsule-based network is extremely difficult due to the added computational cost of storing and routing vector representations, rather than scalars. The original capsule network architecture and dynamic routing algorithm is extremely computationally expensive, both in terms of memory and run-time. Additional intermediate representations are needed to store the output of ``child'' capsules in a given layer while the dynamic routing algorithm determines the coefficients by which these children are routed to the ``parent'' capsules in the next layer. This dynamic routing takes place between every parent and every possible child. One can think of the additional memory space required as a multiplicative increase of the batch size at a given layer by the number of capsule types at that layer. The number of parameters required quickly swells beyond control as well, even for trivially small inputs such as MNIST and CIFAR10. For example, given a set of 32 capsule types with $6 \times 6$, $8$D-capsules per type being routed to $10 \times 1$, $16$D-capsules (as is the case in CapsNet), the number of parameters for this layer alone is $10 \times (6 \times 6 \times 32) \times 16 \times 8 = 1,474,560$ parameters. This one layer contains, coincidentally, roughly the same number of parameters as our entire proposed deep convolutional-deconvolutional capsule network with locally-constrained dynamic routing which itself operates on up to $512 \times 512$ pixel inputs. To scale the original CapsNet up to $512 \times 512$, without these novelties, would require $512\times512\times32\times8\times512\times512\times10\times16 = 2814749767106560$ parameters and hence over $10$ million GB of memory to store the parameters alone.}

\begin{figure*}[t]
  \centering
  \includegraphics[width=\textwidth]{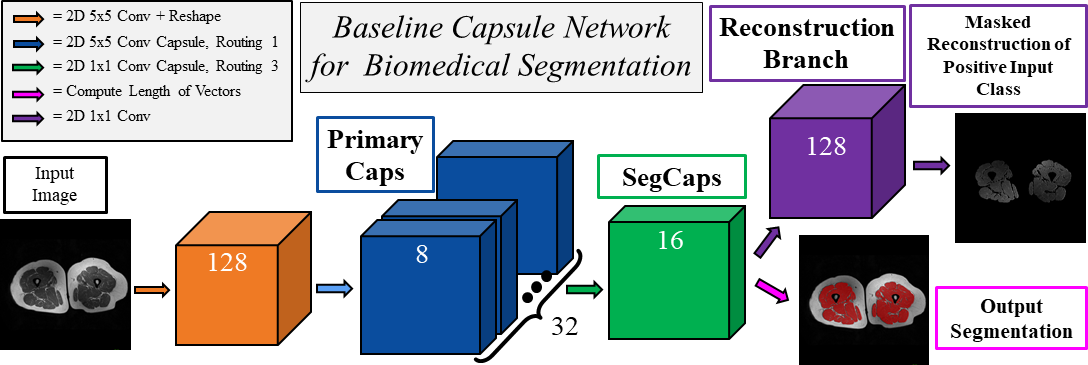}
  \caption{A simple three-layer capsule segmentation network closely mimicking the work by \citet{sabour2017dynamic}. This \ADD{baseline capsule} network uses our proposed locally-constrained dynamic routing algorithm \ADD{with transformation matrix sharing,} as well as the masked reconstruction of the positive input class. \ADD{The input and outputs shown are of muscle tissue segmentation from MRI scans.} }
  \label{fig:CapsSimple}
\end{figure*}

\begin{figure*}[!ht]
  \centering
  \includegraphics[width=\textwidth]{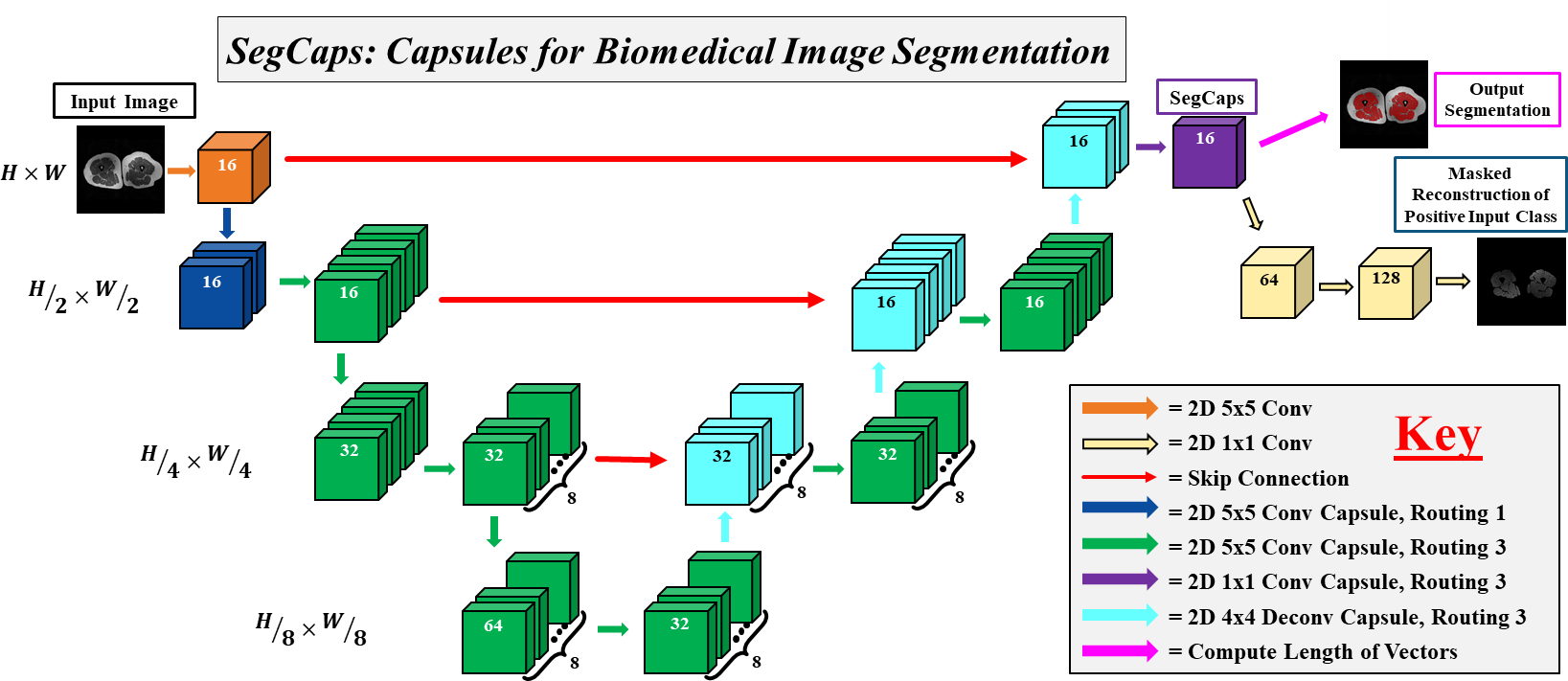}
  \caption{The proposed \textbf{\textit{SegCaps}} architecture for \ADD{biomedical image} segmentation. \ADD{The network is a deep encoder-decoder architecture with skip connections concatenating together capsule types from earlier layer with the same spatial dimensions. The input and outputs shown are from the task of muscle segmentation from MRI scans of patient's thighs.}}
  \label{fig:SegCaps}
\end{figure*}

\ADD{We solve this memory burden and parameter explosion by extending the idea of convolutional capsules (primary capsules in \mbox{\citet{sabour2017dynamic}} are technically convolutional capsules without any routing) and rewriting the dynamic routing algorithm in two key ways. First, children are only routed to parents within a defined spatially-local kernel. Second, transformation matrices are shared for each member of the grid within a capsule type but are not shared across capsule types. To compensate for the loss of global connectivity with the locally-constrained routing, we extend capsule networks by proposing ``deconvolutional'' capsules which operates using transposed convolutions, routed by the proposed locally-constrained routing. These innovations allow us to still learn a diverse set of different capsule types while dramatically reducing the number of parameters in the network, addressing the memory burden. Also, with the proposed deep convolutional-deconvolutional architecture, we retain near-global contextual information and produce state-of-the-art results for our given applications. Our proposed \textbf{\textit{SegCaps}} architecture is illustrated in Figure~\mbox{\ref{fig:SegCaps}}. As a comparative baseline, we also implement a simple three-layer capsule structure, more closely following that of the original capsule implementation, shown in Figure~\mbox{\ref{fig:CapsSimple}}.}

\subsection{\ADD{Summary of Our Contributions}}\label{sec:contrib}

\ADD{The novelty of this paper can be summarized as follows:}
\begin{enumerate}
\item \ADD{Our proposed \textbf{\textit{SegCaps}} is the first use of a capsule network architecture for object segmentation in literature.}
\item \ADD{We propose two technical modifications to the original dynamic routing algorithm where (i) children are only routed to parents within a defined spatially-local window and (ii) transformation matrices are shared for each member of the grid within a capsule type. These modifications, combined with convolutional capsules, allow us to operate on large images sizes (up to $512 \times 512$ pixels) for the first time in literature, where previous capsule architectures typically do not exceed inputs of $32 \times 32$ pixels in size.}
\item \ADD{We introduce the concept of "deconvolutional" capsules and create a novel deep convolutional-deconvolutional capsule architecture, far deeper than the original three-layer capsule network, implement a three-layer convolutional capsule network baseline using our locally-constrained routing to provide a comparison with our \textbf{\textit{SegCaps}} architecture, and extend the masked reconstruction of the target class as a method for regularization to the problem of segmentation as described in Section~\ref{sec:method}.}
\item \ADD{We validate the efficacy of \textbf{\textit{SegCaps}} on the largest-scale study for pathological lung segmentation in the literature, comprising five datasets from both clinical and pre-clinical subjects with nearly $2000$ total CT scans, and $150$ MRI scans at three different contrasts for thigh muscle and adipose (fat) tissue segmentation. For lung segmentation, our proposed method produces improved results in terms of dice coefficient and Hausdorff distance (HD), when compared with state-of-the-art methods \textit{U-Net}~\mbox{\citep{ronneberger2015u}}, \textit{Tiramisu}~\mbox{\citep{jegou2017one}}, and \textit{P-HNN}~\mbox{\citep{harrison2017progressive}}, while dramatically reducing the number of parameters needed to achieve this performance. The proposed \textbf{\textit{SegCaps}} architecture contains $4.6\%$ of the parameters of \textit{U-Net}, $9.5\%$ of \textit{P-HNN}, and $14.9\%$ of \textit{Tiramisu}. Thorough ablation studies are also performed to analyze the contribution and effect of several experimental settings in our proposed model. In particular, there is no other study to conduct fully-automated deep learning based pre-clinical image segmentation due to the extreme levels of variation in both anatomy and pathology present in animal subjects, the large number of high-resolution slices per scan with typically high levels of noise and scanner artifacts, as well as the sheer difficulty in even establishing ground-truth labels compared to human-subject scans.}
\end{enumerate}

\ADD{\subsection{Preliminary Non-Archival SegCaps Study}}

\ADD{In a preliminary non-archival study, we introduced the first-ever capsule-based segmentation network in the literature, which we named \textit{SegCaps}~\mbox{\citep{lalonde2018capsules}}. This initial study demonstrated the ability of SegCaps to perform on-par with state-of-the-art CNNs on the task of pathological lung segmentation from a dataset of CT scans. However, the study was extremely limited in a number of regards: 1) The ground-truth for the dataset was provided by an automated algorithm, raising concerns of similar biases/errors in the proposed algorithm existing within the ground-truth; 2) Results were only presented in terms of Dice scores which do not accurately capture the quality of the segmentation boundaries for the large lung fields; 3) The amount of pathology present in LUNA challenge dataset was fairly limited, targeted at lung nodule detection, rather than segmentation. In this journal extension of our non-archived work, we significantly extend the applications of SegCaps to better demonstrate its generalization abilities and versatility, performing segmentation in both CT and MRI on a wide range of anatomical structures. We obtained radiologist provided annotations for five large-scale pathological lung datasets from both clinical and preclincal subjects covering a wide range of pathologies, as well as dataset of three different MRI contrasts for segmenting muscle and adipose (fat) tissue. We also perform a set of experiments to show a capsule-based segmentation network can better handle changes to viewpoint than a CNN-based approach.}

\section{Background and Related Works}\label{sec:related}

\ADD{Object segmentation in the medical imaging and computer vision communities has remained an interesting and challenging problem over the past several decades. Early attempts in automated object segmentation were analogous to the if-then-else expert systems of that period, where the compound and sequential application of low-level pixel processing and mathematical models were used to build-up complex rule-based systems of analysis \citep{horowitzpicture, Rosenfeld}. In computer vision fields, superpixels and various sets of feature extractors such as scale-invariant feature transform (SIFT) \citep{lowe1999object} or histogram of oriented gradients (HOG) \citep{dalal2005histograms} were used to construct these spaces. Specifically in medical imaging, methods such as level sets \citep{ls}, fuzzy connectedness \citep{fc}, graph-based \citep{gs}, random walk \citep{rw}, and atlas-based algorithms \citep{ab} have been utilized in different application settings. Over time, the community came to favor supervised machine learning techniques, where algorithms were developed using training data to teach systems the optimal decision boundaries in a constructed high-dimensional feature space.}

\ADD{In the last few years, deep learning methods, in particular convolutional neural networks (CNNs), have become the state-of-the-art for various image analysis tasks \citep{ren2015faster, he2016deep, he2017mask, huang2017densely, hu2018squeeze}. Specifically related to the object segmentation problem, \textit{U-Net} \citep{ronneberger2015u}, Fully Convolutional Networks (FCN) \citep{long2015fully}, and other encoder-decoder style CNNs have become the desired models for various medical image segmentation tasks. Most recent attempts in the computer vision and medical imaging literature utilize the extension of these methods to address the segmentation problem \citep{zhao2017pyramid, chen2018encoder, yang2018denseaspp}.}

\subsection{\ADD{CNN-Based Segmentation}}\label{sec:related_dlseg}

 The object segmentation literature is vast, both before and in the deep learning era. Herein, we only summarize the most popular deep learning-based segmentation algorithms. Based on FCN \citep{long2015fully} for semantic segmentation, \textit{U-Net} \citep{ronneberger2015u} introduced an alternative CNN-based pixel label prediction algorithm which forms the backbone of many deep learning-based segmentation methods in medical imaging today. Following this, many subsequent works follow this encoder-decoder structure, experimenting with dense connections, skip connections, residual blocks, and other types of architectural additions to improve segmentation accuracy for particular imaging applications. For instance, a recent example by \citet{jegou2017one} combines a \textit{U-Net}-like structure with the very successful DenseNet \citep{huang2017densely}  architecture, creating a densely connected \textit{U-Net} structure, called \textit{Tiramisu}. Other successful frameworks for segmentation and their specific innovations are the following. 

SegNet \citep{badrinarayanan2017segnet} attempts to improve the upsampling process by performing ``unpooling'', capturing the pooling indices from the max pooling layers in the encoder to more accurately place features in the decoder feature maps. Although the encoder-decoder structure is specifically designed to capture global context information, several methods attempt to further improve this global context in different ways. RefineNet \citep{lin2017refinenet} fuses features from multiple resolutions through adding residual connections and chained residual pooling to create a large cascaded encoder-decoder structure. PSPNet \citep{zhao2017pyramid} introduces a pyramid pooling module by pooling at different kernel sizes and  concatenating back to the features maps. Large Kernel Matters \citep{peng2017large} uses large $1 \times 15 + 15 \times 1$ and $15 \times 1 + 1 \times 15$ global convolution networks. ClusterNet \citep{lalonde2018clusternet} combines two fully-convolutional networks, one to capture global and one for local information, to segment specifically a large number of densely packed tiny objects, normally lost in networks with pooling. DeepLab \citep{chen2018deeplab} utilizes an atrous spatial pyramid pooling (ASPP) unit to better capture image context from multiple scales. The latest version of DeepLab (v3+) \citep{chen2018encoder} follows a very similar structure to \textit{U-Net} with the addition of an ASPP for image context and depthwise separable convolutions for efficiency.

\ADD{\subsection{Segmentation in Biomedical Imaging}} \label{sec:related_biomed}

\ADD{As mentioned in Section~\ref{sec:intro}, segmentation is of critical importance as a first stage in many biomedical imaging applications. Though well motivated, performing segmentation within biomedical imaging introduces a number of unique challenges, including handling many different imaging modalities anatomical structures, and potential deformities/abnormalities caused by a wide range of reasons. Further, imaging data across different applications can be 2D, 3D, and even 4D, requiring unique considerations. Multi-view networks, such as \citep{mortazi2017multi}, remain a popular approach to handling imaging data with more than two dimensions. 3D networks such as 3D U-Net~\mbox{\citep{cciccek20163d}} and V-Net~\mbox{\citep{milletari2016v}} have also gained recent popularity based off the highly successful U-Net~\mbox{\citep{ronneberger2015u}}. Nonetheless, due to a combination of limited GPU memory and the desire to exploit existing pretrained models, majority of the literature uses 2D network and analyzes 3D data in a slice-wise manner. In this study, we focus on two of the most commonly investigated imaging modalities, namely CT and MRI, and detail the specific related works to those applications in the following paragraphs.}

\subsection{Pathological Lung Segmentation from CT}\label{sec:related_lungseg}

Anatomy and pathology segmentation have been central to the most medical imaging applications. Despite its importance, accurate segmentation of pathological lungs from CT scans remains extremely challenging due to a wide spectrum of lung abnormalities such as consolidations, ground glass opacities, fibrosis, honeycombing, tree-in-buds, and nodules. \ADD{Specifically developed for pathological lung segmentation,~\mbox{\citet{mansoor2014generic}} created a two-stage approach based on fuzzy connectedness and texture features, incorporating anatomical information by segmenting the rib-cage. Most recently,~\mbox{\citet{harrison2017progressive}} developed \textit{P-HNN}, which achieved very strong results on a subset of three clinical datasets by modifying the Holistically-Nested Network (HNN)~\mbox{\citep{xie15hed}} structure to progressively sum side-output predictions during the decoder phase.} In this study, we test the efficacy of the proposed \textbf{\textit{SegCaps}} algorithm for pathological lung segmentation due to precise segmentation's importance as a precursor to the deployment of nearly any computer-aided diagnosis (CAD) tool for pulmonary image analysis.

\subsection{\ADD{Muscle and Adipose (Fat) Tissue Segmentation from MRI}}\label{sec:related_mri}

\ADD{A number of applications favor MRI as the primary imaging modality, including most popularly cardiac applications. For example,~\mbox{\citet{mortazi2017cardiacnet}} proposed a multi-view CNN, following an encoder-decoder structure and adding a novel loss function, for segmenting the left atrium and proximal pulmonary veins from MRI. Body composition analysis (\textit{e.g.} segmenting/quantifying muscle and adipose (fat) tissue) favors MRI as well, due to its excellent soft tissue contrast and lack of ionizing radiation. In \mbox{\citet{irmakci2018novel}}, the authors proposed a method based on fuzzy connectivity to perform segmentation of muscle and fat tissue of the thigh region of whole-body MRI scans. This work represents the current state of the art results in terms of Dice score. }

\subsection{\ADD{Capsule Networks}}\label{sec:related_caps}

\ADD{A simple three-layer capsule network, called \textit{CapsNet}, showed remarkable initial results in~\mbox{\citet{sabour2017dynamic}}, producing state-of-the-art classification results on the MNIST dataset and relatively good classification results on the CIFAR10 dataset. Since then, researchers have begun extending the idea of capsule networks to other applications, including brain-tumor classification~\mbox{\citep{afshar2018brain}}, lung-nodule screening~\mbox{\citep{mobiny2018fast}}, action detection~\mbox{\citep{duarte2018videocapsulenet}}, point-cloud autoencoders~\mbox{\citep{zhao20193d}}, adversarial detection~\mbox{\citep{frosst2018darccc, qin2019detecting}}, and even creating wardrobes~\mbox{\citep{hsiao2018creating}}, as well as several technical contributes to improve the routing mechanism for datasets such as MNIST, CIFAR10, SVHN, SmallNorb, etc.~\mbox{\citep{hinton2018matrix, kosiorek2019stacked}}. Nonetheless, the majority of these works remain focused on small image classification, and no work yet exists in literature for a method of capsule-based object segmentation.}

The remainder of the paper is organized as follows: Section~\ref{sec:method} describes our proposed \textit{SegCaps} framework in detail, including the deconvolutional capsules and reconstruction regularization for segmentation; Section~\ref{sec:results} details the five experimental datasets, our implementation settings (e.g. hyperparameters), and the results of our main experiments; Section~\ref{sec:ablations} covers the ablation studies performed which help to determine the contribution of each aspect of our proposed method to the final results; and finally Section~\ref{sec:conclusion} is the discussion and conclusions of our work.
Experimental results of our method applied to other types of imaging data and applications to provide empirical support for the general applicability of our study are included in the appendix.

\section{\textit{SegCaps}: Capsules for Object Segmentation} \label{sec:method}

In the following section, we describe the formulation of our \textit{SegCaps} architecture. As illustrated in Figure 2, the input to our \textit{SegCaps} network is a large image (e.g. $512 \times 512$ pixels), in this case, a slice of a  \ADD{MRI} Scan. The image is passed through a $2$D convolutional layer which produces $16$ feature maps of the same spatial dimensions. This output forms our first set of capsules, where we have a single capsule type with a grid of $512 \times 512$ capsules, each of which is a $16$ dimensional vector. This is then followed by our first convolutional capsule layer. In the following, we generalize the process of our convolutional capsules and routing to any given layer $\ell$ in the network. 

At layer $\ell$, there exists a set of capsule types 
\begin{equation}\label{eq:child_caps_types}
T^\ell = \{t_1^\ell, t_2^\ell, ..., t_n^\ell \mid n \in \mathbb{N}\}. 
\end{equation}
For every $t_i^\ell \in T^\ell$, there exists an $h^\ell \times w^\ell$ grid of $z^\ell$-dimensional child capsules, 
\begin{equation}\label{eq:child_caps}
C = \{\bm{c}_{1, 1}, ..., \bm{c}_{1, w^\ell}, ..., \bm{c}_{h^\ell, 1}, ..., \bm{c}_{h^\ell, w^\ell}\}, 
\end{equation}
where $h^\ell \times w^\ell$ is the spatial dimensions of the output of layer $\ell-1$. At the next layer of the network, $\ell+1$, there exists a set of capsule types 
\begin{equation}\label{eq:parent_caps_types}
T^{\ell+1} = \{t_1^{\ell+1}, t_2^{\ell+1}, ..., t_m^{\ell+1} \mid m \in \mathbb{N}\}. 
\end{equation}
And for every $t_j^{\ell+1} \in T^{\ell+1}$, there exists an $h^{\ell+1} \times w^{\ell+1}$ grid of $z^{\ell+1}$-dimensional parent capsules, 
\begin{equation}\label{eq:parent_caps}
P = \{\bm{p}_{1, 1}, ..., \bm{p}_{1, w^{\ell+1}}, ..., \bm{p}_{h^{\ell+1}, 1}, ..., \bm{p}_{h^{\ell+1}, w^{\ell+1}}\},
\end{equation}
where $h^{\ell+1} \times w^{\ell+1}$ is the spatial dimensions of the output of layer $\ell$.

In convolutional capsules, for every parent capsule type $t_j^{\ell+1} \in T^{\ell+1}$, every parent capsule $\bm{p}_{x, y} \in P$ receives a set of ``prediction vectors'', $\{\bm{\hat{u}}_{x, y \mid t_1^\ell}, \bm{\hat{u}}_{x, y \mid t_2^\ell}, ..., \bm{\hat{u}}_{x, y \mid t_n^\ell}\}$, one for each capsule type in $T^\ell$. This set of prediction vectors is defined as the matrix multiplication between a learned transformation matrix for the given parent capsule type, $M_{t_j^{\ell+1}}$, and the sub-grid of child capsules outputs, $U_{x, y \mid t_i^\ell}$, within a user-defined kernel centered at position $(x, y)$ in layer $\ell$; hence 
\begin{equation}\label{eq:pred_vectors}
\bm{\hat{u}}_{x, y \mid t_i^\ell} = M_{t_j^{\ell+1}} \cdot U_{x, y \mid t_i^\ell}, \quad \forall \enspace  t_i^\ell \in T^\ell.
\end{equation}
Explicitly, each $U_{x, y \mid t_i^\ell}$ has shape $k_h \times k_w \times z^\ell$, where $k_h \times k_w$ are the dimensions of the user-defined kernel, for all capsule types $t_i^\ell \in T^\ell$. Each $M_{t_j^{\ell+1}}$ has shape $k_h \times k_w \times z^\ell \times z^{\ell+1}$. Thus, we can see each $\bm{\hat{u}}_{x, y \mid t_i^\ell}$ is an $z^{\ell+1}$-dimensional vector, since these will be used to form our parent capsules. In practice, we solve for all parent capsule types simultaneously by defining $M$ to have shape $k_h \times k_w \times z^\ell \times \mid T^{\ell+1} \mid \times z^{\ell+1}$, where $\mid T^{\ell+1} \mid$ is the number of parent capsule types in layer $\ell+1$. Note, as opposed to CapsNet, we are sharing transformation matrices across members of the grid (i.e. each $M_{t_j^{\ell+1}}$ does not depend on the spatial location $(x,y)$), as the same transformation matrix is shared across all spatial locations within a given capsule type, similar to how convolutional kernels scan an input feature map. This is one way our method can exploit parameter sharing to dramatically cut down on the total number of parameters to be learned. The values of these transformation matrices for each capsule type in a layer are learned via the backpropagation algorithm with a supervised loss function.

\begin{algorithm*}
\caption{Locally-Constrained Dynamic Routing.}\label{routingalg}
\begin{algorithmic}[1]
\Procedure{Routing}{$\bm{\hat{u}}_{x, y\mid t_i^\ell}$, $d$, $\ell$, $x$, $y$}
\State for all capsule types $t_i^\ell$ at position $(x,y)$ and capsule type $t_j^{\ell+1}$ at position $(x, y)$: $b_{t_i^\ell \mid x, y} \gets 0$.
\For{$d$ iterations}
\State for all capsule types $t_i^\ell$ at position $(x,y)$: ${\bf r}_{t_i^\ell} \gets \texttt{softmax}({\bf b}_{t_i^\ell})$ \Comment{\texttt{softmax} computes Eq.~\ref{eq:softmax}}
\State for all capsule types $t_j^{\ell+1}$ at position $(x, y)$: $\bm{p}_{x, y} \gets \sum_n r_{t_i^\ell \mid x, y} \bm{\hat{u}}_{x, y \mid t_i^\ell}$
\State for all capsule types $t_j^{\ell+1}$ at position $(x, y)$: $\bm{v}_{x, y} \gets \texttt{squash}(\bm{p}_{x, y})$ \Comment{\texttt{squash} computes Eq.~\ref{eq:squash}}
\State for all capsule types $t_i^\ell$ and all capsule types $t_j^{\ell+1}$: $ b_{t_i^\ell \mid x, y} \gets b_{t_i^\ell \mid x, y} + \bm{\hat{u}}_{x, y \mid t_i^\ell} . {\bf v}_{x, y}$
\EndFor
\Return ${\bf v}_{x, y}$
\EndProcedure
\end{algorithmic}
\end{algorithm*}

To determine the final input to each parent capsule $\bm{p}_{x, y} \in P$, where again $P$ is the grid of parent capsules for parent capsule type $t_j^{\ell+1} \in T^{\ell+1}$, we compute the weighted sum over these ``prediction vectors'' as,
\begin{equation}\label{eq:routing}
\bm{p}_{x, y} = \sum_n r_{t_i^\ell \mid x, y} \bm{\hat{u}}_{x, y \mid t_i^\ell}, 
\end{equation}
where $r_{t_i^\ell \mid x, y}$ are the routing coefficients determined by the dynamic routing algorithm, and each member of the grid $(x, y)$ has a unique routing coefficient. These routing coefficients are computed by a ``routing softmax'', 
\begin{equation}\label{eq:softmax}
r_{t_i^\ell \mid x, y} = \frac{\exp(b_{t_i^\ell \mid x, y})}{\sum_{t_j^{\ell+1}} \exp(b_{t_i^\ell \mid t_j^{\ell+1}})}, 
\end{equation}
whose initial logits, $b_{t_i^\ell \mid x, y}$ are the log prior probabilities that prediction vector $\bm{\hat{u}}_{x, y\mid t_i^\ell}$ should be routed to parent capsule $\bm{p}_{x, y}$. Note that the $\sum_{t_j^{\ell+1}}$ term is across parent capsule types in $T^{\ell+1}$ for each $(x, y)$ location.

Our method differs from the dynamic routing implemented by \citet{sabour2017dynamic} in two ways. First, we locally constrain the creation of the prediction vectors. Second, we only route the child capsules within the user-defined kernel to the parent, rather than routing every single child capsule to every single parent. The output capsule is then computed using a non-linear squashing function 
\begin{equation}\label{eq:squash}
{\bf v}_{x, y} = \frac{||\bm{p}_{x, y}||^2}{1+||\bm{p}_{x, y}||^2} \frac{\bm{p}_{x, y}}{||\bm{p}_{x, y}||},
\end{equation}
where ${\bf v}_{x, y}$ is the vector output of the capsule at spatial location $(x,y)$ and $\bm{p}_{x, y}$ is its final input. Lastly, the agreement is measured as the scalar product, 
\begin{equation}\label{eq:activation}
a_{x, y \mid t_i^\ell} = {\bf v}_{x, y} \cdot \bm{\hat{u}}_{x, y\mid t_i^\ell}.
\end{equation}
The pseudocode for this locally-constrained dynamic routing is summarized in Algorithm~\ref{routingalg}. A final segmentation mask is created by computing the length of the capsule vectors in the final layer and assigning the positive class to those whose magnitude is above a threshold, and the negative class otherwise. 

\subsection{Deconvolutional Capsules}\label{sec:deconv}

In order to form a deep encoder-decoder network, we introduce the concept of ``deconvolutional'' capsules. These are similar to the locally-constrained convolutional capsules; however, the prediction vectors are now formed using the transpose of the operation previously described. Note that the dynamic routing of these differently-formed prediction vectors still occurs in the exact same way, so we will not re-describe that part of the operation. 

The set of prediction vectors for deconvolutional capsules are defined again as the matrix multiplication between a learned transformation matrix, $M_{t_j^{\ell+1}}$, for a given parent capsule type $t_j^{\ell+1} \in T^{\ell+1}$, and the sub-grid of child capsules outputs, $W_{x, y \mid t_i^\ell}$ for each capsule type in $t_i^\ell \in T^\ell$, within a user-defined kernel centered at position $(x, y)$ in layer $\ell$. However, in deconvolutional capsules, we first need to reshape our child capsule outputs following the fractional striding formulation used in \citet{long2015fully}. This allows us to effectively upsample the height and width of our capsule grids by the scaling factor chosen. For each member of the grid, we can then form our prediction vectors again by 
\begin{equation}\label{eq:deconv_pred_vectors}
\bm{\hat{w}}_{x, y \mid t_i^\ell} = M_{t_j^{\ell+1}} \cdot W_{x, y \mid t_i^\ell}, \quad \forall \enspace  t_i^\ell \in T^\ell.
\end{equation}
Thus, we have each $\bm{\hat{w}}_{x, y \mid t_i^\ell}$ as a $z^{\ell+1}$-dimensional vector, and is input to the dynamic routing algorithm to form our parent capsules. As before, in practice we solve for all parent capsule types simultaneously by defining $M$ to have shape $k_h \times k_w \times z^\ell \times \mid T^{\ell+1} \mid \times z^{\ell+1}$, where $\mid T^{\ell+1} \mid$ is the number of parent capsule types in layer $\ell+1$. Here, we still sharing transformation matrices across members of the grid (i.e. each $M_{t_j^{\ell+1}}$ does not depend on the spatial location $(x,y)$), similar to how transposed convolutional kernels scan an input feature map.

\begin{figure*}[ht]
  \centering
  \includegraphics[width=1\textwidth]{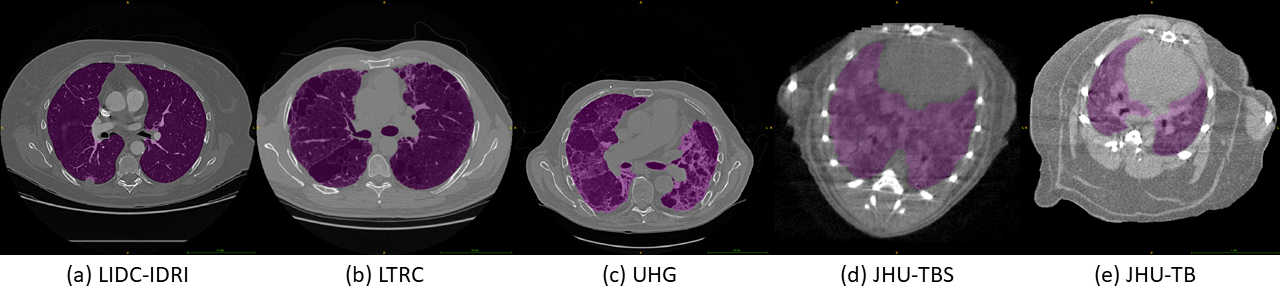}
  \caption{Example scans with ground-truth masks (magenta) for each of the five datasets in this study.}
  \label{fig:datasets}
\end{figure*}

\subsection{Reconstruction Regularization}\label{sec:recon}

As a method of regularization, we extend the idea of reconstructing the input to promote a better embedding of our input space. This forces the network to not only retain all necessary information about a given input, but also encourages the network to better represent the full distribution of the input space, rather than focusing only on its most prominent modes relevant to the desired task. Since we only wish to model the distribution of the positive input class and treat all other pixels as background, we mask out segmentation capsules which do not belong to the positive class and reconstruct a similarly masked version of the input image. We perform this reconstruction via a three layer $1 \times 1$ convolutional network, then compute a mean-squared error (MSE) loss between only the positive input pixels and this reconstruction. More explicitly, we formulate this problem as
\begin{equation}\label{eq:maskedimg}
R^{x,y} = I^{x,y}\times S^{x,y} \mid S^{x,y} \in \{0,1\}\mbox{, and}
\end{equation}
\begin{equation}\label{eq:maskedloss}
\mathcal{L}_R = \frac{\gamma}{X \times Y}\sum_x^X\sum_y^Y\|R^{x,y} - O_r^{x,y}\|,
\end{equation}
where $\mathcal{L}_R$ is the supervised loss for the reconstruction regularization, $\gamma$ is a weighting coefficient for the reconstruction loss, $R^{x,y}$ is the reconstruction target pixel, $I^{x,y}$ is the image pixel, $S^{x,y}$ is the ground-truth segmentation mask value, and $O_r^{x,y}$ is the output of the reconstruction network, each at pixel location $(x,y)$, respectively, and $X$ and $Y$ are the width and height, respectively, of the input image. \ADD{For simplicity $\gamma$ was initially set to $1$; however, the parameter produces similar results for settings from $1$ -- $0.001$. Lower than this or higher than this starts to degrade performance,} An ablation study of the contribution of this regularization is included in Section~\ref{sec:ablations}. The total loss is the summation of this reconstruction loss and a weighted binary cross-entropy (BCE) loss for the segmentation output, weighted by the foreground/background  pixel  balance  of  each  training  set respectively.

\section{Experiments and Results}\label{sec:exps}

\subsection{Pathological Lung Datasets}\label{sec:data}

Experiments were conducted on five pathological lung datasets, obtained from both clinical and pre-clinical subjects, containing nearly $2000$ CT scans, with annotations by expert radiologists. An example typical scan with ground-truth from each dataset is shown in Figure~\ref{fig:datasets}. The three clinical and two pre-clinical (mice) datasets analyzed are as follows: 
\begin{itemize}
    \item \textit{The Lung Image Database Consortium and Image Database Resource Initiative} \citep{lidc}, abbreviated as \textbf{LIDC-IDRI}, contains $885$ annotated CT scans of lung cancer screening patients collected from seven academic centers and eight medical imaging companies. \ADD{Scans were captured using seven different GE Medical Systems LightSpeed scanner models, four different Philips Brilliance scanner models, five different Siemens Deﬁnition, Emotion, and Sensation scanner models, and Toshiba Aquilion scanners. Slice thicknesses range from $0.6$ to $5$ mm with in-plane pixel sizes ranging from $0.461$ to $0.977$ mm. All image sizes are $512 \times 512$ pixels per slice.}
    \item \textit{The Lung Tissue Research Consortium database} \citep{ltrc}, abbreviated as \textbf{LTRC}, contains $545$ annotated CT scans, with most donor subjects having interstitial fibrotic lung disease or chronic obstructive pulmonary disease (COPD). \ADD{All  scans using the LTRC protocol were obtained using either General Electric or Siemens scanners with 16 or more detectors, and imaging parameters were standardized as much as possible among the enrollment centers (with slice thickness $1.25$ mm or less with $50$\% overlapping reconstruction in a high-spatial, frequency-preserving algorithm). All image sizes are $512 \times 512$ pixels per slice.}
    \item \textit{The Multimedia Database of Interstitial Lung Diseases} \citep{ild}, abbreviated as \textbf{UHG}, built at the University Hospitals of Geneva contains $214$ annotated CT scans of patients affected with one of the $13$ histological diagnoses of interstitial lung disease (ILD). \ADD{The dataset follows the HRCT scanning protocol, with a slice thickness of $1$ -– $2$ mm, spacing between slices of $10$ -– $15$ mm, scan time of $1$ -– $2$ s, no contrast agent, axial pixel matrix of $512 \times 512$, and $x$, $y$ spacing of $0.4$ -– $1$ mm.}
    \item \textit{The TB-Smoking dataset} collected at Johns Hopkins University, abbreviated as \textbf{JHU-TBS}, contains $108$ annotated CT scans of mice subjects affected with tuberculosis (TB) and exposed to smoke inhalation. \ADD{Slice thicknesses range from $0.1$ to $0.2$ mm with in-plane pixel sizes ranging from $0.1$ to $0.2$ mm. Images range in size from $176 \times 176$ to $352 \times 352$ pixels per slice.}
    \item \textit{The TB dataset} also collected at Johns Hopkins University, abbreviated as \textbf{JHU-TB}, contains $208$ annotated CT scans of mice subjects affected with TB undergoing experimental treatment. \ADD{Slice thicknesses range from $0.041$ to $0.058$ mm with in-plane pixel sizes ranging from $0.041$ to $0.058$ mm. Images range in size from $199 \times 212$ to $580 \times 496$ pixels per slice.}
\end{itemize}

In total, $1960$ CT scans were annotated in this study. Each dataset was treated completely separate, as each offers unique challenges to automated segmentation algorithms. \ADD{For preprocessing, all CT scans were clipped at $-1024$ and $3072$, then normalized to a $0-1$ scale. All images used their original resolutions during training and testing.} Ten-fold cross-validation was performed for training all algorithms, with $10\%$ of training data left aside for validation and early-stopping. The mean and standard deviation (std) across the 10-folds for each dataset is presented for  \ADD{two} key metrics, namely the 3D Dice similarity coefficient (Dice) and 3D Hausdorff distance (HD) computer for each 3D CT scan.

\subsection{Implementation Details}\label{sec:implement}

All algorithms, namely \textit{U-Net}, \textit{Tiramisu}, \textit{P-HNN}, our three-layer baseline capsule segmentation network, and \textbf{\textit{SegCaps}} are all implemented using Keras \citep{chollet2015keras} with TensorFlow \citep{tensorflow}. The \textit{U-Net} architecture is implemented exactly as described in the original paper by \citet{ronneberger2015u}. \textit{P-HNN} was implemented based on their official Caffe code, including individual layer-specific learning rate multipliers and kernel initialization. However, we removed the layer-specific learning rate and changed the kernel initialization to Xavier to match the other networks and achieve much better results. \textit{Tiramisu} follows the highest performing model presented in \citep{jegou2017one}, namely \textit{FC-DenseNet103}. To remain consistent, since pre-trained models are not available for our custom-designed \textit{SegCaps}, and to better see the performance of each individual method under different amounts of training data and pathologies present, no pre-trained weights were used to initialize any of the models; instead, all were trained from scratch on each dataset investigated. It can be reasonably assumed based on previous studies that pre-training on large datasets such as ImageNet would improve the performance of all models. A weighted-BCE loss is used for the segmentation output of all networks, with weights determined by the foreground/background pixel balance of each training set respectively. For the capsule network, the reconstruction output loss is computed via the masked-MSE described in Section~\ref{sec:method}. All possible experimental factors are controlled between different networks; all networks are trained from scratch, using the same data augmentation methods (scale, flip, shift, rotate, elastic deformations, and random noise) and Adam optimization \citep{kingma2014adam} with an initial learning rate of $0.00001$. A batch size of $1$ is chosen for all experiments to match the original \textit{U-Net} implementation. The learning rate is decayed by a factor of $0.05$ upon validation loss stagnation for $50,000$ iterations and early-stopping is performed with a patience of $250,000$ iterations based on validation 2D Dice scores. Positive/negative pixels were set in the segmentation masks based on a threshold of on the networks' output score maps. Thresholds are found dynamically for each testing split based on which level provides the best Dice score for the validation set of images. All code is made publicly available.~\footnote{\href{https://github.com/lalonderodney/SegCaps}{https://github.com/lalonderodney/SegCaps}}


\subsection{Lung Segmentation Results}\label{sec:results}

The final quantitative results of these experiments to perform lung segmentation from pathological CT scans are shown in Tables~\ref{table:lidc} -~\ref{table:jhu-tb}. Table~\ref{table:lidc} shows results on the LIDC-IDRI dataset, the largest of the three clinical datasets with typically the least severe pathology present on average compared to the other two clinical datasets. Table~\ref{table:ltrc} shows results on the LTRC dataset, a large dataset with large amounts of ILD and COPD pathology present. Table~\ref{table:uhg} shows results on the UHG dataset, perhaps the most challenging of the three clinical datasets, both due to its relatively smaller size and the severe average amount of pathology present in patients scanned. Table~\ref{table:jhu-tbs} shows results on the JHU-TBS dataset, and provides the first fully-automated deep learning based segmentation results presented in the literature for lung segmentation on pre-clinical subjects. Table~\ref{table:jhu-tb} shows results on the JHU-TB dataset, a larger but more challenging dataset of mouse subjects with typically more severe pathology present than the JHU-TBS dataset.

\begin{table}[h!]
\centering
\caption{Experimental results on 885 CT scans from the LIDC-IDRI database \citep{lidc}, measured by 3D Dice Similarity Coefficient and Hausdorff Distance (HD).}
\label{table:lidc}
\resizebox{\linewidth}{!}{\begin{tabular}{@{}c|cc@{}}
\toprule
Method & Dice ($\% \pm$ std) & HD ($mm \pm$ std) \\ 
\midrule
\textit{U-Net} \citep{ronneberger2015u} & $96.06 \pm 2.40$ & $41.211 \pm 9.109$ \\
\textit{Tiramisu} \citep{jegou2017one} & $94.40 \pm 3.66$ & $42.205 \pm 15.210$ \\
\textit{P-HNN} \citep{harrison2017progressive} & $95.64 \pm 2.92$ & $41.775 \pm 13.866$ \\
\textbf{\textit{SegCaps}} & $\bm{96.98 \pm 0.36}$ & $\bm{30.764 \pm 2.793}$ \\
\bottomrule
\end{tabular}}
\end{table}

\begin{table}[h!]
\centering
\caption{Experimental results on 545 CT scans from the LTRC database \citep{ltrc}, measured by 3D Dice Similarity Coefficient and Hausdorff Distance (HD).}
\label{table:ltrc}
\resizebox{\linewidth}{!}{\begin{tabular}{@{}c|cc@{}}
\toprule
Method & Dice ($\% \pm$ std) & HD ($mm \pm$ std) \\ 
\midrule
\textit{U-Net} \citep{ronneberger2015u} & $95.52 \pm 2.80$ & $37.625 \pm 6.831$ \\
\textit{Tiramisu} \citep{jegou2017one} & $95.41 \pm 2.08$ & $43.969 \pm 14.869$ \\
\textit{P-HNN} \citep{harrison2017progressive} & $95.46 \pm 3.93$ & $33.835 \pm 9.596$ \\
\textbf{\textit{SegCaps}} & $\bm{96.91 \pm 2.24}$ & $\bm{26.295 \pm 3.806}$ \\
\bottomrule
\end{tabular}}
\end{table}

\begin{table}[h!]
\centering
\caption{Experimental results on 214 CT scans from the UHG database \citep{ild}, measured by 3D Dice Similarity Coefficient and Hausdorff Distance (HD).}
\label{table:uhg}
\resizebox{\linewidth}{!}{\begin{tabular}{@{}c|cc@{}}
\toprule
Method & Dice ($\% \pm$ std) & HD ($mm \pm$ std) \\ 
\midrule
\textit{U-Net} \citep{ronneberger2015u} & $88.10 \pm 1.84$ & $44.303 \pm 34.148$ \\
\textit{Tiramisu} \citep{jegou2017one} & $87.67 \pm 1.38$ & $61.227 \pm 54.096$ \\
\textit{P-HNN} \citep{harrison2017progressive} & $88.64 \pm 0.64$ & $43.698 \pm 24.026$ \\
\textbf{\textit{SegCaps}} & $\bm{88.92 \pm 0.66}$ & $\bm{37.171 \pm 23.223}$ \\
\bottomrule
\end{tabular}}
\end{table}

\begin{table}[h!]
\centering
\caption{Experimental results on 108 CT scans from the JHU-TBS database, measured by 3D Dice Similarity Coefficient and Hausdorff Distance (HD).}
\label{table:jhu-tbs}
\resizebox{\linewidth}{!}{\begin{tabular}{@{}c|ccc@{}}
\toprule
Method & Dice ($\% \pm$ std) & HD ($mm \pm$ std) \\ 
\midrule
\textit{U-Net} \citep{ronneberger2015u} & $90.38 \pm 3.86$ & $7.593 \pm 0.886$ \\
\textit{Tiramisu} \citep{jegou2017one} & $86.45 \pm 5.76$ & $7.428 \pm 1.337$ \\
\textit{P-HNN} \citep{harrison2017progressive} & $88.81 \pm 6.81$ & $7.517 \pm 1.896$ \\
\textbf{\textit{SegCaps}} & $\bm{93.35 \pm 0.95}$ & $\bm{4.367 \pm 1.367}$ \\
\bottomrule
\end{tabular}}
\end{table}

\begin{table}[h!]
\centering
\caption{Experimental results on 208 CT scans from the JHU-TB database, measured by 3D Dice Similarity Coefficient and Hausdorff Distance (HD).}
\label{table:jhu-tb}
\resizebox{\linewidth}{!}{\begin{tabular}{@{}c|ccc@{}}
\toprule
Method & Dice ($\% \pm$ std) & HD ($mm \pm$ std) \\ 
\midrule
\textit{U-Net} \citep{ronneberger2015u} & $76.26 \pm 9.51$ & $\bm{24.295 \pm 14.684}$ \\
\textit{Tiramisu} \citep{jegou2017one} & $79.99 \pm 6.24$ & $24.647 \pm 11.629$ \\
\textit{P-HNN} \citep{harrison2017progressive} & $80.11 \pm 7.46$ & $26.597 \pm 16.168$ \\
\textbf{\textit{SegCaps}} & $\bm{80.91 \pm 5.27}$ & $26.021 \pm 10.260$ \\
\bottomrule
\end{tabular}}
\end{table}

The results of these experiments show \textbf{\textit{SegCaps}} consistently outperforms all other compared state-of-the-art approaches in terms of the commonly measured metrics, Dice and HD. Additionally, \textit{SegCaps} achieves this while only using a fraction of the total parameters of these much larger networks. The proposed \textbf{\textit{SegCaps}} architecture contains  \ADD{less than $4.6$\%} parameters than \textit{U-Net},  \ADD{less} than \ADD{$9.5$ \% of} \textit{P-HNN}, and  \ADD{less} than \ADD{$14.9$ \% of} \textit{Tiramisu}. A comparison with similarly sized version of these other networks is shown in Section~\ref{sec:abl_params}. As a brief note in regardless to the discrepancy in results for \textit{P-HNN} between our study and those in the original work, this can be explained by several factors: the original work i) used ImageNet pre-trained models, ii) selected a carefully chosen subset (73 scans) of the UHG dataset, , and iii) trained and tested models using all datasets combined in the cross-validation splits.

\begin{figure*}[h!]
  \centering
  \includegraphics[width=0.7\textwidth]{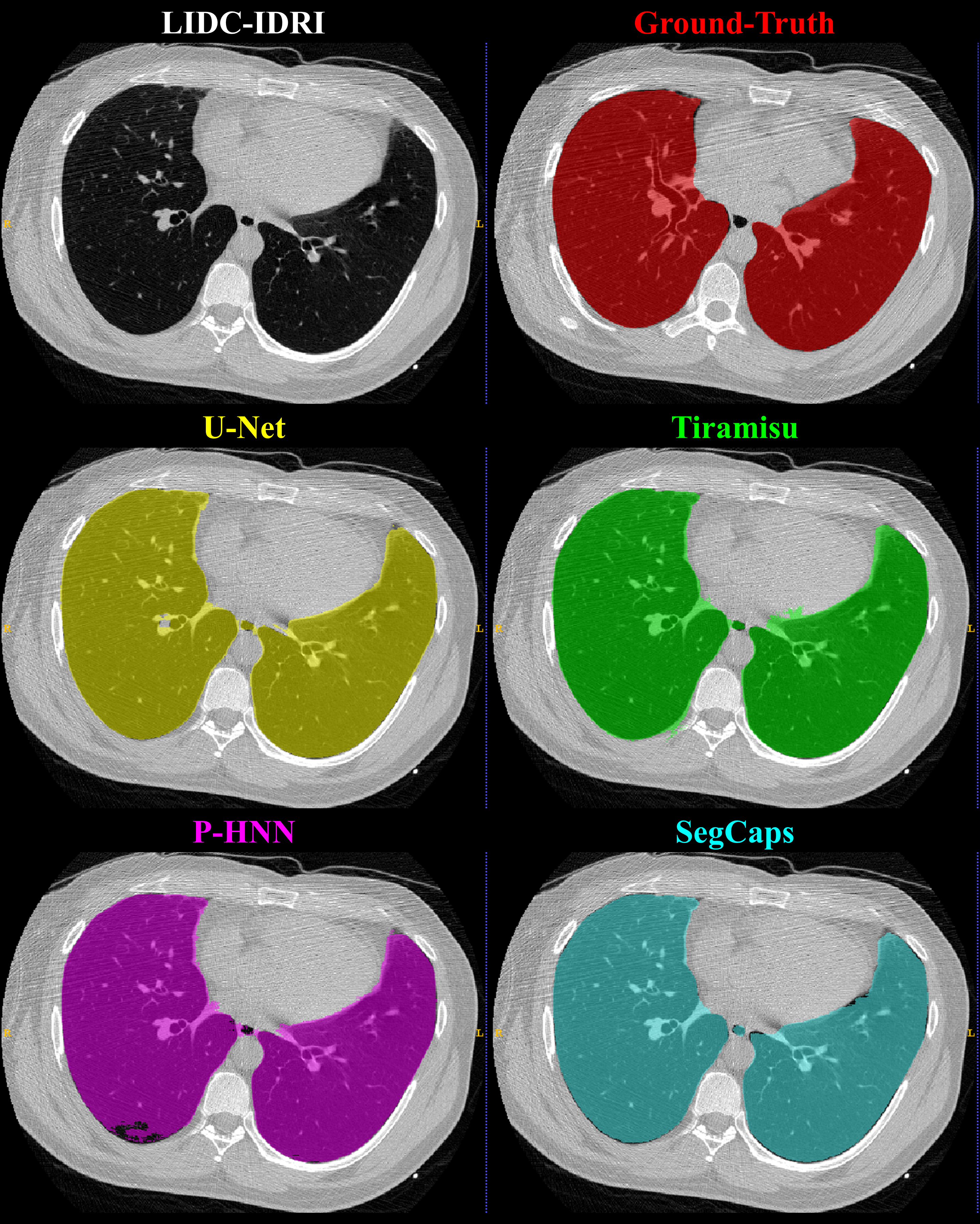}
  \caption{\ADD{Qualitative results for a 2D slice from a CT scan taken from the LIDC-IDRI dataset. It can be noticed that the CNN-based methods' typical failure cases are where the pixel intensities (Hounsfield units) are far from the class mean (\textit{i.e.} high values within the lung regions or low values outside the lung regions).}}
  \label{fig:LIDC_results}
\end{figure*}

\begin{figure*}[h!]
  \centering
  \includegraphics[width=0.7\textwidth]{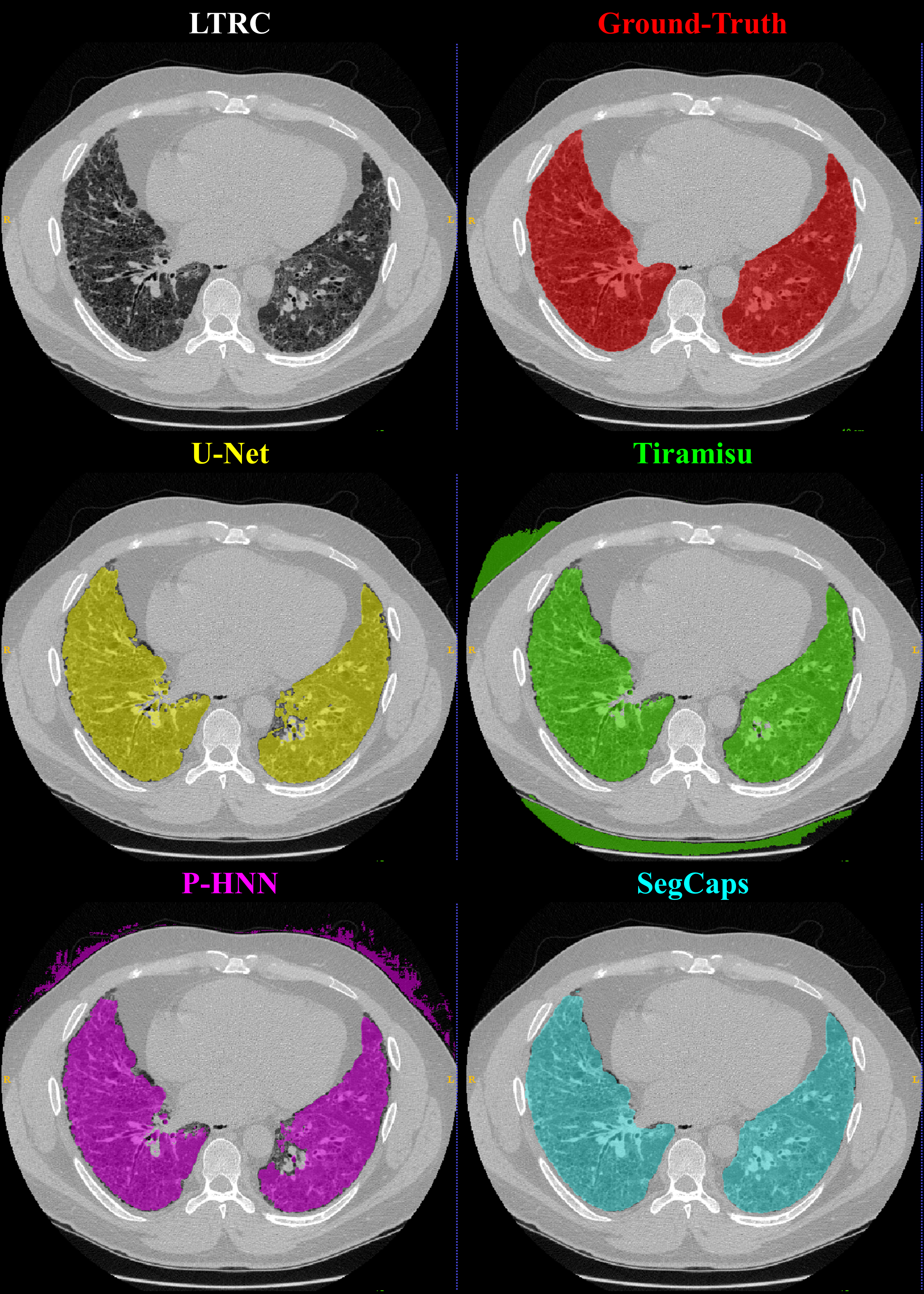}
  \caption{\ADD{Qualitative results for a 2D slice from a CT scan taken from the LTRC dataset. It can be noticed that the CNN-based methods' typical failure cases are where the pixel intensities (Hounsfield units) are far from the class mean (\textit{i.e.} high values within the lung regions or low values outside the lung regions).}}
  \label{fig:LTRC_results}
\end{figure*}

\begin{figure*}[h!]
  \centering
  \includegraphics[width=0.7\textwidth]{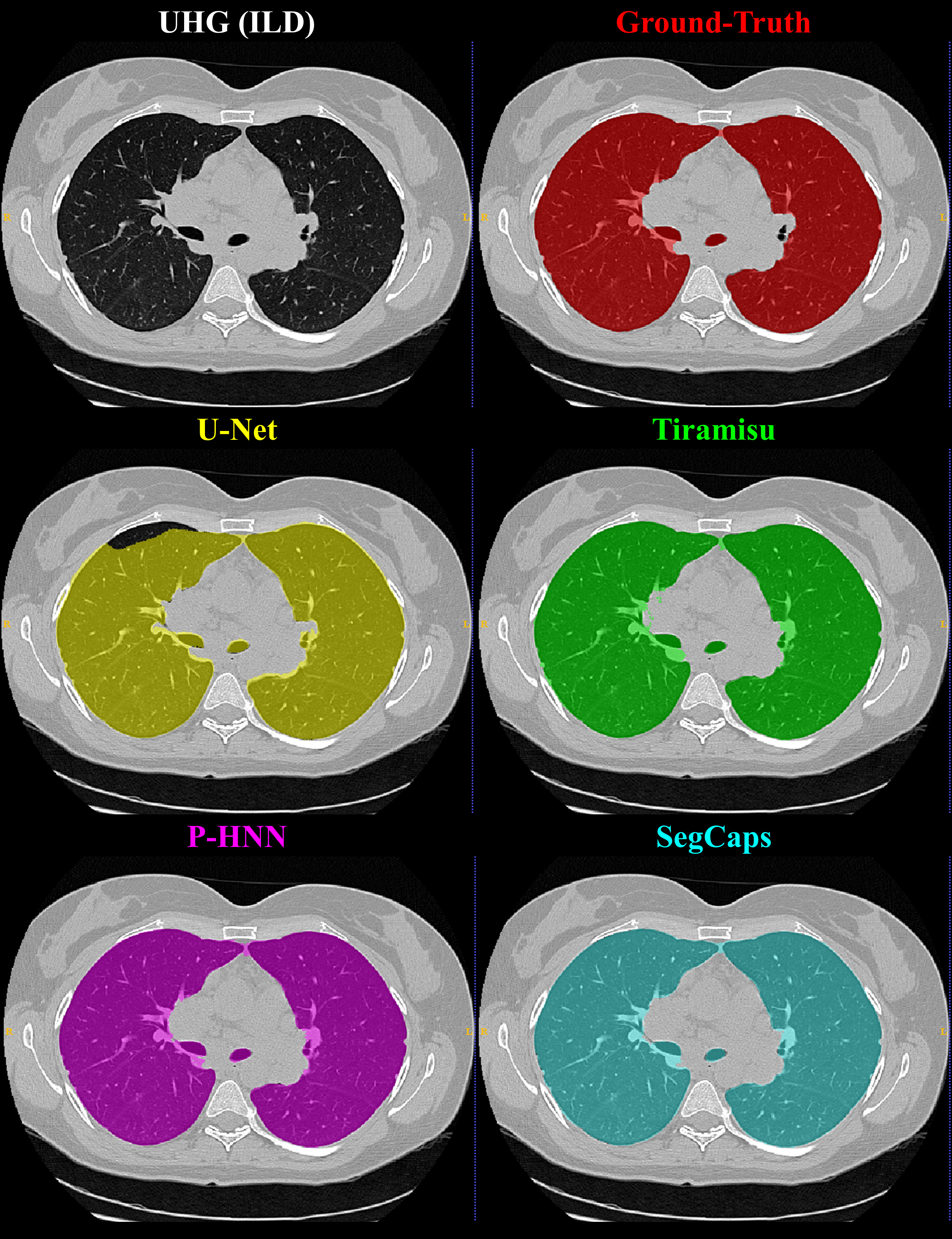}
  \caption{\ADD{Qualitative results for a 2D slice from a CT scan taken from the UHG dataset. It can be noticed that the CNN-based methods' typical failure cases are where the pixel intensities (Hounsfield units) are far from the class mean (\textit{i.e.} high values within the lung regions or low values outside the lung regions).}}
  \label{fig:ILD_results}
\end{figure*}

\begin{figure*}[h!]
  \centering
  \includegraphics[width=0.7\textwidth]{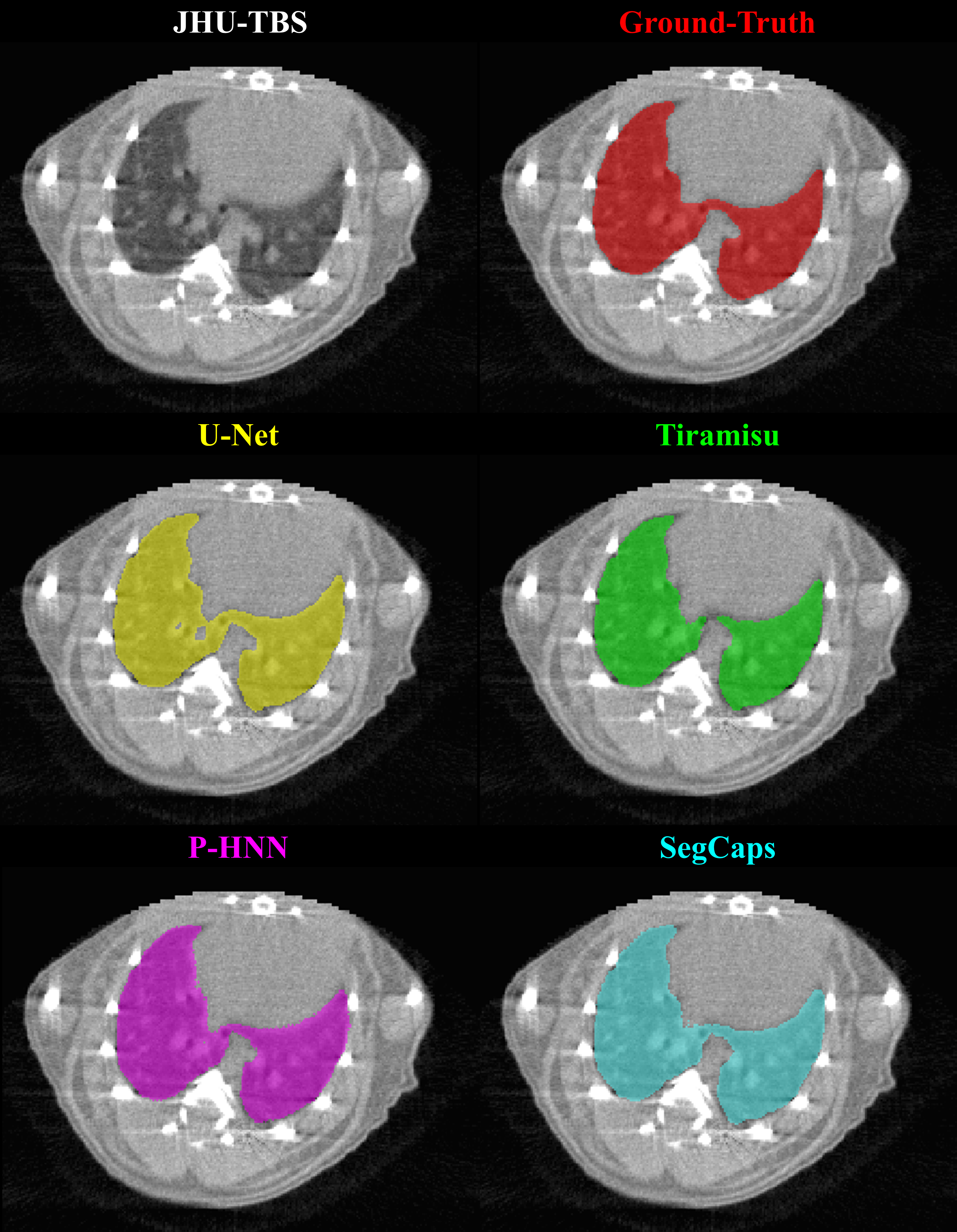}
  \caption{\ADD{Qualitative results for a 2D slice from a CT scan taken from the JHU-TBS dataset. Note this drastically different anatomy and high level of noise present in the preclinical mice subjects. It can be noticed that the CNN-based methods' typical failure cases are where the pixel intensities (Hounsfield units) are far from the class mean (\textit{i.e.} high values within the lung regions or low values outside the lung regions).}}
  \label{fig:TBS_results}
\end{figure*}

\begin{figure}[h!]
  \centering
  \includegraphics[width=0.45\textwidth]{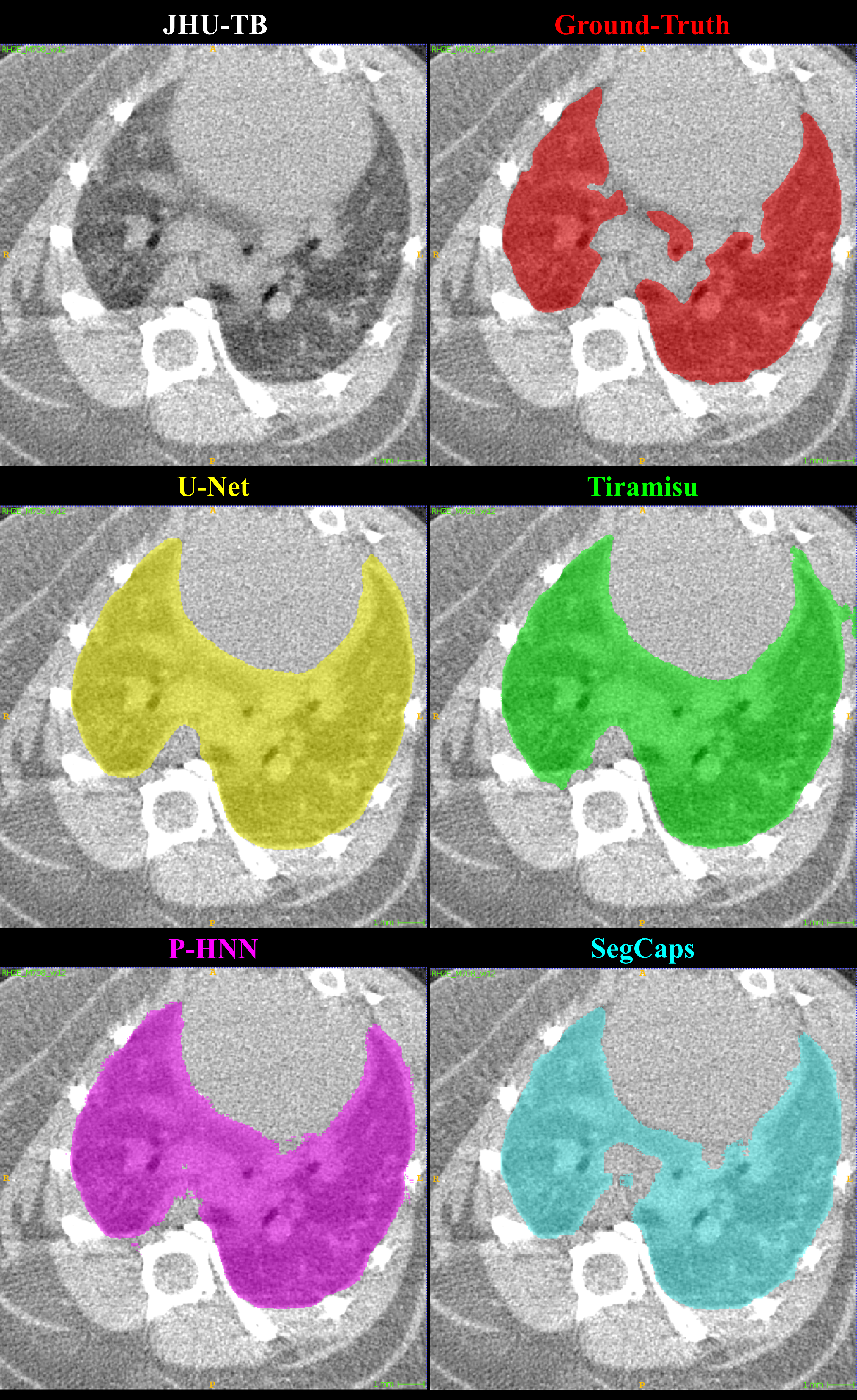}
  \caption{\ADD{Qualitative results for a 2D slice from a CT scan taken from the JHU-TB dataset. Note this drastically different anatomy and high level of noise present in the preclinical mice subjects. It can be noticed that the CNN-based methods' typical failure cases are where the pixel intensities (Hounsfield units) are far from the class mean (\textit{i.e.} high values within the lung regions or low values outside the lung regions).}}
  \label{fig:DRTB_results}
\end{figure}

Qualitative results for typical samples from all datasets are shown in Figure\ADD{s}~\ref{fig:LIDC_results} -- \ADD{\ref{fig:DRTB_results}}. As can be seen in these qualitative examples, \textit{SegCaps} achieves higher results by not falling into the typical segmentation failure causes, namely over-segmentation and segmentation-leakage. These qualitative examples are supported by our quantitative findings where over-segmentation is best captured by the HD metric and segmentation-leakages are best captured by the Dice metric. 

\begin{figure}[h!]
  \centering
  \includegraphics[width=0.45\textwidth]{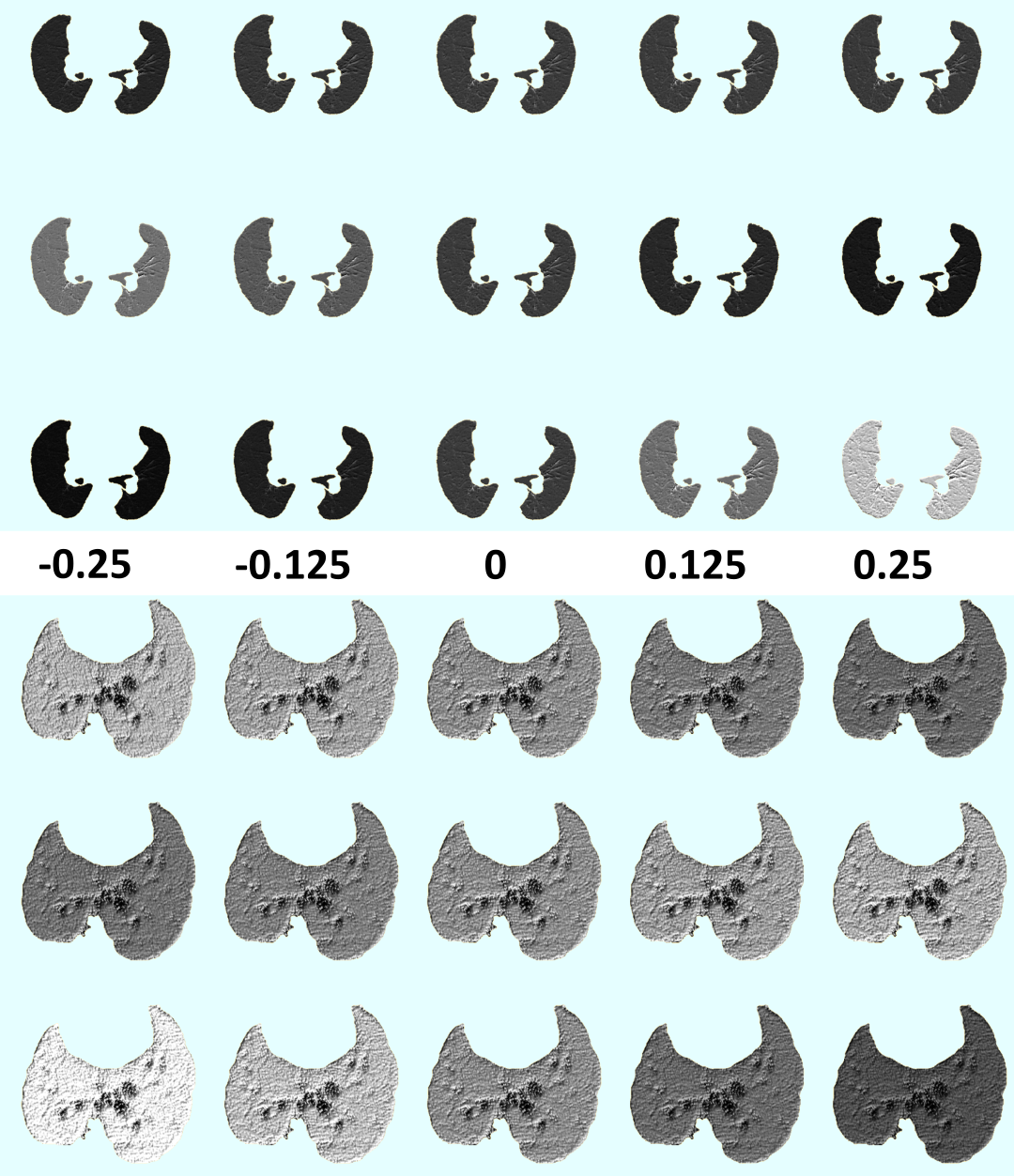}
  \caption{Reconstructions of selected capsule vectors (rows) under different perturbations from $-0.25$ -- $0.25$ (columns). The top three rows are reconstructions of a scan slice from the clinical LTRC dataset, while the bottom three are from the pre-clinical JHU-TB dataset. These results demonstrate that different dimensions of the capsule vectors are in fact learning different attributes of the lung tissue being segmented.}
  \label{fig:manip}
\end{figure}

Further, we investigate how different capsule vectors in the final segmentation capsule layer are representing different visual attributes. Figure~\ref{fig:manip} shows three selected visual attributes (each row) out of the sixteen (dimension of final capsule segmentation vector) across different perturbation values of the vectors ranging from -$0.25$ to +$0.25$ (each column) for an example clinical and pre-clinical scan. We observe that regions with different textural properties (i.e., small and large homogeneous) are progressively captured by the different dimensions of the capsule segmentation vectors.

\ADD{\subsection{Muscle and Adipose (Fat) Tissue Segmentation Datasets and Preprocessing}} \label{sec:fat_datasets}

\begin{figure*}[h!]
  \centering
  \includegraphics[width=0.7\textwidth]{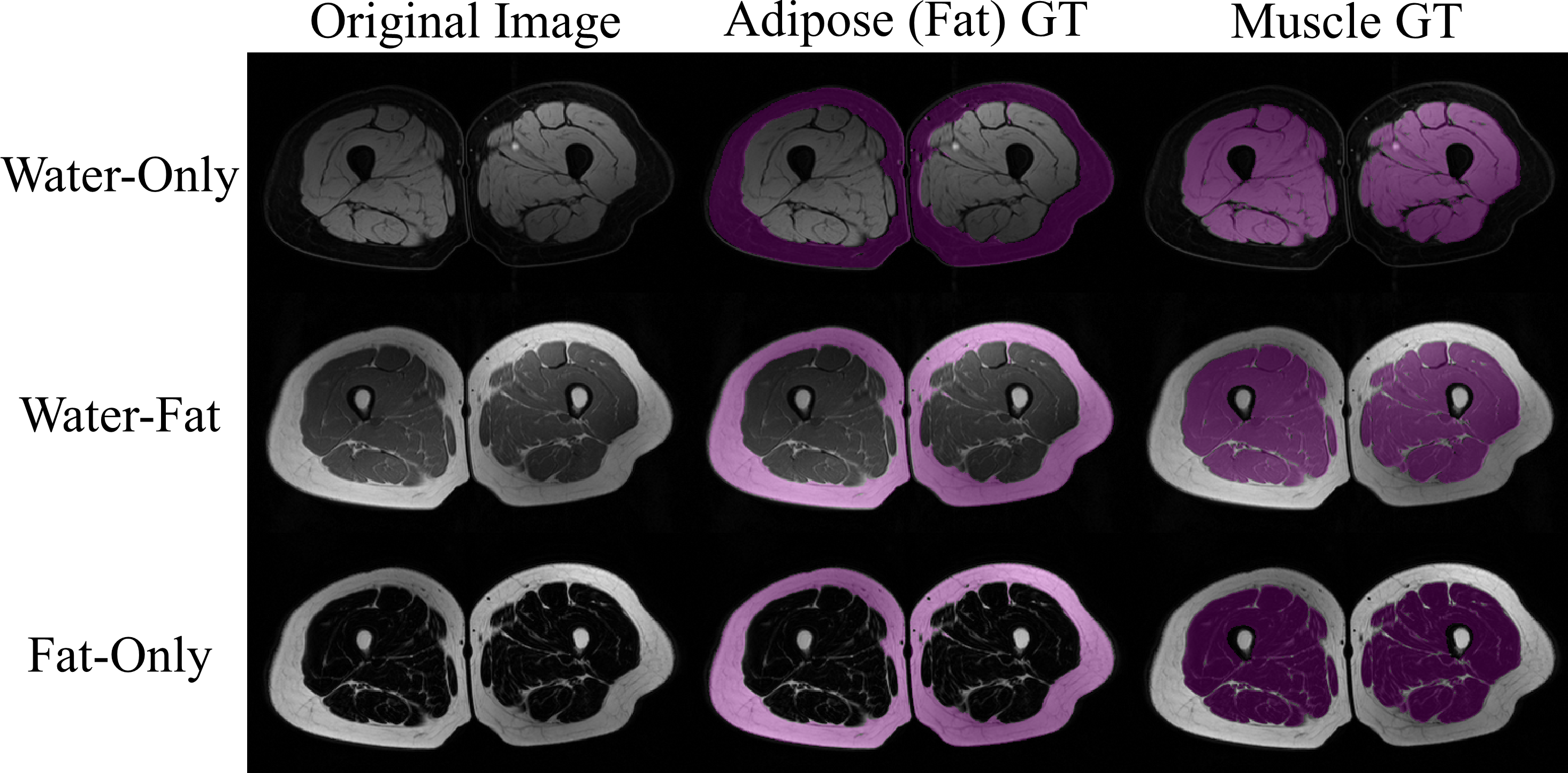}
  \caption{\ADD{Example magnetic resonance (MR) images from the Baltimore Longitudinal Study of Ageing (BLSA)~\mbox{\citep{ferrucci2008baltimore}} dataset. Three different T1-weighted MR contrasts, namely water and fat, water-only (fat-suppressed), and fat-only (water-suppressed) are shown with their ground-truth (GT) annotations.}}
  \label{fig:MRI_data}
\end{figure*}

\ADD{Experiments were conducted on the Baltimore Longitudinal Study of Ageing (BLSA)~\mbox{\citep{ferrucci2008baltimore}}, where a total of 150 scans were collected using three contrasts from 50 subjects. These MRI were acquired using a 3T Philips Achieva MRI scanner (Philips Healthcare, Best, The Netherlands) equipped with a Q-body radiofrequency coil for transmission and reception. Three different T1-weighted MR contrasts, namely water and fat, water-only (fat-suppressed), and fat-only (water-suppressed), were used, where water and fat suppression were achieved using spectral pre-saturation with inversion recovery (SPIR), with coverage from the proximal to distal ends of the femur using $80$ slices in the foot to head direction, a field of view (FOV) of $440 \times 296 \times 400$ mm$^3$ and a voxel size of $1 \times 1$ mm$^2$ in-plane, and slice thickness varies from 1 mm to 3 mm in different scans (one particular scan was with 5 mm slice thickness). The age of subjects ranged between $44-89$ years and the body mass index (BMI) ranged from $18.67 - 45.68$. Examples of each MRI contrast with the ground truth-annotations (GT) for both muscle and adipose (fat) tissue are shown in Figure~\ref{fig:MRI_data}.}

\ADD{For training and testing, we performed preprocessing on the MRI images. First we applied the  non-uniform non-parametric intensity normalisation technique by~\mbox{\citet{tustison2010n4itk}} to
remove field bias. Next we apply Curvature Anisotropic Diffusion to smooth the image and remove noise. Lastly we perform z-score normalization to normalize the intensities before standardizing the image to the $0-1$ range.}

\ADD{\subsection{Muscle and Adipose (Fat) Tissue Segmentation Results}} \label{sec:fat_results}

\begin{figure*}[h!]
  \centering
  \includegraphics[width=\textwidth]{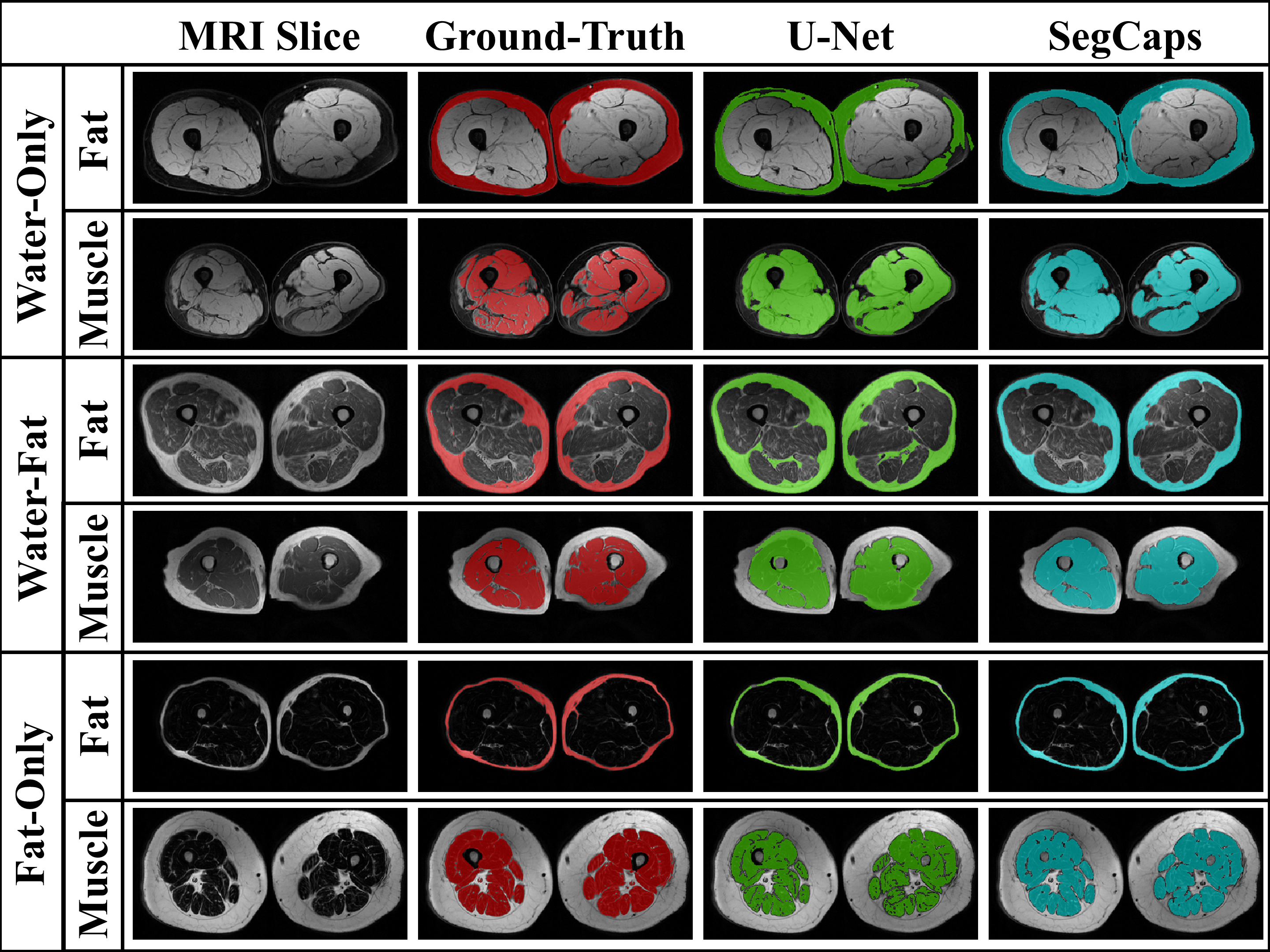}
  \caption{\ADD{Qualitative results on the BLSA dataset for T1-weighted water-only (fat-suppressed), water-fat, and fat-only (water-suppressed) contrasts of six different patients. Results are shown for U-Net and SegCaps, with U-Net showing systematic issues with areas of similar intensities values to foreground class, but which actually belong to the background class. and vice-versa.}}
  \label{fig:MRI_results}
\end{figure*}

\ADD{Experiments were performed using U-Net and SegCaps and compared to the state-of-the-art method by~\mbox{\citet{irmakci2018novel}} using the same comparative metric (\textit{i.e.} Dice coefficient). The results of these experiments are shown in Tables~\ref{table:fat_1} -- ~\ref{table:fat_3} and show that both U-Net and SegCaps can outperform the previous state-of-the-art, while SegCaps does so using again only a small fraction of the parameters as U-Net while performing at the same level as U-Net. The results reported for~\mbox{\citet{irmakci2018novel}} are the results from the original work (using manual seeding) on the same dataset. For U-Net and SegCaps, qualitative results are shown in Fig~\ref{fig:MRI_results} for six of the $50$ patients in the BLSA dataset. As with the lung segmentation experiments, U-Net tends to struggle with areas of similar intensity values, which do not belong to the correct tissue class.}

\begin{table}[h!]
\centering
\caption{\ADD{Experimental results on water-only (fat-suppressed) MRI scans from the BLSA dataset, measured by 3D Dice Similarity Coefficient.}}
\label{table:fat_1}
\resizebox{\linewidth}{!}{\begin{tabular}{@{}c|cc@{}}
\toprule
\ADD{Method} & \ADD{Adipose (Fat) ($\% \pm$ std)} & \ADD{Muscle ($\% \pm$ std)} \\ 
\midrule
\ADD{\textit{FC}~\mbox{\citep{irmakci2018novel}}} & \ADD{$44.46 \pm 27.29$} & \ADD{$67.70 \pm 24.67$} \\
\ADD{\textit{U-Net}~\mbox{\citep{ronneberger2015u}}} & \ADD{$\bm{84.48} \pm 8.33$} & \ADD{$90.00 \pm 1.85$} \\
\ADD{\textbf{\textit{SegCaps}}} & \ADD{$84.45 \bm{\pm 6.60}$} & \ADD{$\bm{90.74 \pm 1.49}$} \\
\bottomrule
\end{tabular}}
\end{table}

\begin{table}[h!]
\centering
\caption{\ADD{Experimental results on water-fat MRI scans from the BLSA dataset, measured by 3D Dice Similarity Coefficient.}}
\label{table:fat_2}
\resizebox{\linewidth}{!}{\begin{tabular}{@{}c|cc@{}}
\toprule
\ADD{Method} & \ADD{Adipose (Fat) ($\% \pm$ std)} & \ADD{Muscle ($\% \pm$ std)} \\ 
\midrule
\ADD{\textit{FC}~\mbox{\citep{irmakci2018novel}}} & \ADD{$78.52 \pm 14.77$} & \ADD{$84.56 \pm 14.66$} \\
\ADD{\textit{U-Net}~\mbox{\citep{ronneberger2015u}}} & \ADD{$91.11 \pm 3.10$} & \ADD{$90.34 \pm 8.76$} \\
\ADD{\textbf{\textit{SegCaps}}} & \ADD{$\bm{91.26 \pm 2.77}$} & \ADD{$\bm{92.59 \pm 1.14}$} \\
\bottomrule
\end{tabular}}
\end{table}

\begin{table}[h!]
\centering
\caption{\ADD{Experimental results on fat-only (water suppressed) MRI scans from the BLSA dataset, measured by 3D Dice Similarity Coefficient.}}
\label{table:fat_3}
\resizebox{\linewidth}{!}{\begin{tabular}{@{}c|cc@{}}
\toprule
\ADD{Method} & \ADD{Adipose (Fat) ($\% \pm$ std)} & \ADD{Muscle ($\% \pm$ std)} \\ 
\midrule
\ADD{\textit{FC}~\mbox{\citep{irmakci2018novel}}} & \ADD{$77.52 \pm 16.38$} & \ADD{$73.00 \pm 17.78$} \\
\ADD{\textit{U-Net}~\mbox{\citep{ronneberger2015u}}} & \ADD{$\bm{94.61} \pm 2.67$} & \ADD{$\bm{93.85 \pm 1.01}$} \\
\ADD{\textbf{\textit{SegCaps}}} & \ADD{$94.28 \bm{\pm 2.59}$} & \ADD{$93.38 \pm 1.42$} \\
\bottomrule
\end{tabular}}
\end{table}

\ADD{\subsection{Generalizing to Unseen Orientations of Objects}} \label{sec:single_img}

\begin{figure*}[h!]
  \centering
  \includegraphics[width=0.92\textwidth]{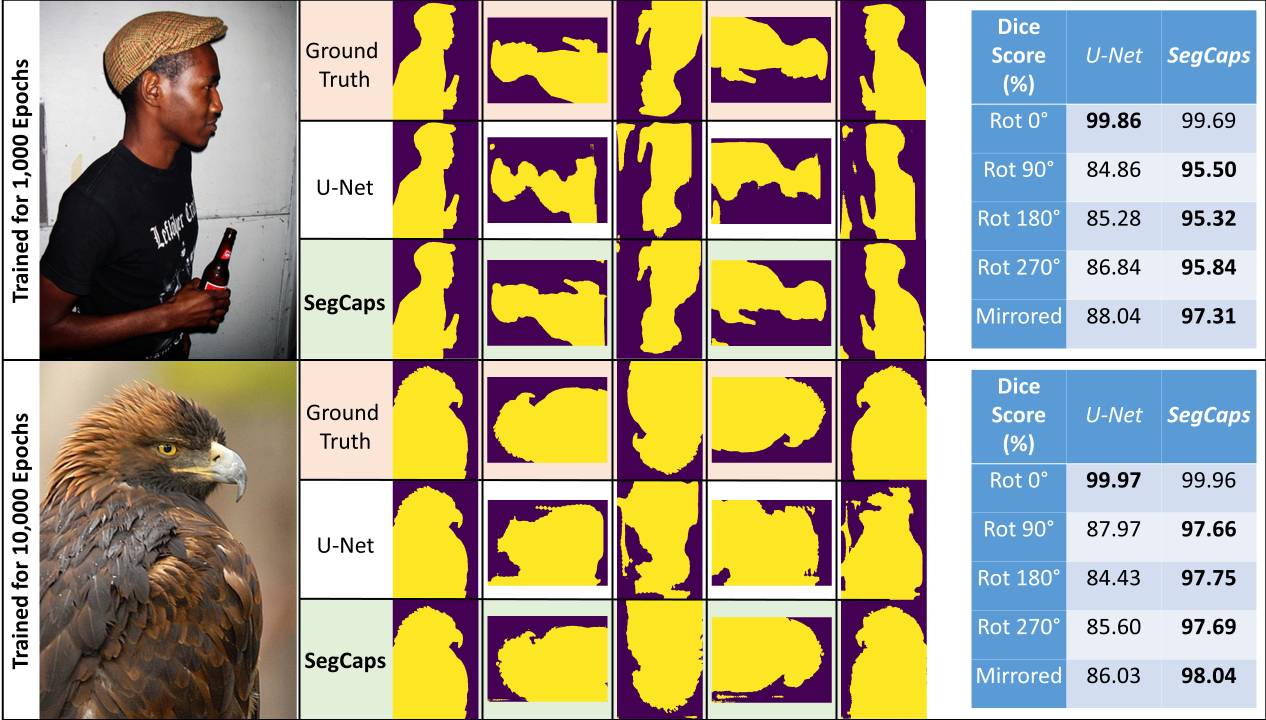}
  \caption{Testing the affine equivariant properties of capsule networks, specifically \textit{SegCaps}, by overfitting on a single image, trained without augmentation, then predicting on transformations of that image.}
  \label{fig:single_img}
\end{figure*}

\ADD{In a final set of experiments, we tested the affine equivariant property of capsule networks on natural images. It has been stated that, due to the affine projections of capsule vectors from children to parents, capsules should be robust to affine transformations on the input, and should in fact be able to generalize to \textit{unseen} poses of target classes. However, no study has formally demonstrated this property. In this experiment, we randomly selected images from the PASCAL VOC dataset which contained only a single foreground object. Both \textit{U-Net} and \textit{SegCaps} were then trained on a single selected image until training accuracy converge to $100\%$, which occurred around $1000$ epochs for both networks. For training, each network followed exactly the training settings described in Section~\ref{sec:implement}. Each network was then tested on $90$ degree rotations and the mirroring of the training image. \textit{SegCaps} performed well on nearly all images tested, while \textit{U-Net} performed quite poorly, as can be seen in Figure~\ref{fig:single_img}. Since \textit{U-Net} has significantly more parameters than \textit{SegCaps}, we also ran experiments at $10000$ epochs, long after both networks had converged to $100\%$ training accuracy. \textit{U-Net} continued to present failures, where \textit{SegCaps} did not suffer the same issue. This shows that \textit{SegCaps} is indeed far more robust to affine transformations on the input, a significant issue for CNNs as shown in both this experiment and works such as by~\mbox{\citet{alcorn2019strike}}.}

\section{Ablations Studies}\label{sec:ablations}

In the following subsections, we investigate the role of the deeper encoder-decoder network structure enabled by the introduction of our deconvolutional capsules, the effect of the reconstruction regularization, the optimal number of dynamic routing iterations to perform, and the relative efficiency of parameter use with similarly-sized versions of all studied networks. The UHG dataset is perhaps the most challenging of the three clinical \ADD{lung segmentation} datasets in our study, both due to its relatively smaller size and the average amount of pathology present in patients scanned. As seen in Table~\ref{table:uhg}, results on all metrics are significantly lower for this challenging dataset. For those reasons, and the lower performance scores leading to bigger differences between approaches, as well as the dataset being publicly available, we chose this dataset for running our ablation experiments.

\subsection{Network Structure/Deconvolutional Capsules}\label{sec:abl_deconv}

The original CapsNet introduced by \citet{sabour2017dynamic} was a simple three layer network, consisting of a single convolutional layer, a primary capsule layer (convolutional layer with a reshape function), and a fully-connected capsule layer. This network achieved remarkable results for its size, beating the state-of-the-art on MNIST and performing well on CIFAR10. In our initial efforts for this study, we attempted to apply this network to the task of segmentation, however, the fully-connected capsule layer was far too memory intensive to make this approach viable with our $512 \times 512$ 2D slices of CT scans. After introducing the locally-constrained dynamic routing and transformation matrix sharing, we then created a network nearly identical to the original CapsNet with the fully-connected capsule layer swapped out for our locally-constrained version. A diagram of this network is shown in Figure~\ref{fig:CapsSimple}. The results of this network on the UHG dataset is shown in Table~\ref{table:baseline}. As one might expect, swapping out a layer which is fully-connected in space for one which is locally-connected dramatically hurt the performance for a task which relies on global information (\textit{i.e.} determining lung tissue/air from non-lung tissue, bone, etc.). This motivated the introduction of the ``deconvolutional'' capsule layer which allows for the creation of deep encoder-decoder networks, and thus the recovery of global information, retention of local information, and the parameter savings of locally-constrained capsules. 

\begin{table}[h!]
\centering
\caption{Comparing the deeper encoder-decoder network structure \textit{SegCaps} enabled by our proposed deconvolutional capsules, versus a network designed to be as similar as possible to CapsNet~\citep{sabour2017dynamic} (\textit{Baseline SegCaps}), abbreviated in table as Base-Caps.}
\label{table:baseline}
\begin{tabular}{@{}c|cc@{}}
\toprule
Method & Dice ($\% \pm$ std) & HD ($mm \pm$ std) \\ 
\midrule
Base-Caps & $75.97 \pm 4.60$ & $352.582 \pm 133.451$ \\
\textit{SegCaps} & $\bm{88.92 \pm 0.66}$ & $\bm{37.171 \pm 23.223}$ \\
\bottomrule
\end{tabular}
\end{table}

\subsection{Parameter Use}\label{sec:abl_params}

\begin{table}[h!]
\centering
\caption{Number of parameters for each of the networks examined in this study. The percentage of less parameters (Percent Less) is measured relative to the number of parameters in \textit{U-Net}.}
\label{table:params}
\begin{tabular}{@{}c|cr@{}}
\toprule
Method & Parameters & Percent Less \\ 
\midrule
\textit{U-Net} & $31.0$ M &  \ADD{Baseline ($100$ \%)} \\
\textit{P-HNN} & $14.7$ M &  \ADD{$47.42$} \% \\
\textit{Tiramisu} & $9.4$ M &  \ADD{$30.32$} \% \\
\textit{Baseline SegCaps} & $1.7$ M &  \ADD{$5.48$} \% \\
\textbf{\textit{SegCaps}} & $\bm{1.4}$ \textbf{M} &  \ADD{$\bm{4.52}$} \textbf{\%} \\
\bottomrule
\end{tabular}
\end{table}

Shown in Tables~\ref{table:params}--~\ref{table:downscaled}, we investigate the number of parameters in the proposed \textit{SegCaps}, \textit{U-Net}, \textit{Tiramisu} and \textit{P-HNN}, as well as down-scaled versions of \textit{U-Net}, \textit{Tiramisu}, and \textit{P-HNN}. \textit{U-Net} and \textit{P-HNN} are scaled down by dividing the number of feature maps per layer by a constant factor, $k=4.68$ and $k=3.2$ respectively, and \textit{Tiramisu} is scaled down by using the lighter FC-DenseNet56 purposed in the original work by \citet{jegou2017one}. When the parameters of \textit{U-Net} and \textit{P-HNN} are scaled down to roughly the same number of parameters as \textit{SegCaps}, these models perform comparatively worse, as shown in Table~\ref{table:downscaled}, providing evidence that \textit{SegCaps} is able to make better use of the parameters available to it than its CNN counterparts. \textit{Tiramisu-56} is a minor exception to this trend as its Dice score remained similar while the HD only fell slightly from \textit{Tiramisu-103}. The reason for this is most likely because \textit{Tiramisu-56} was carefully engineered to achieve the highest possible accuracy with few parameters while the addition of dense connections has been shown to make far better use of parameters than standard non-dense CNNs \citep{huang2017densely}. However, as can be see in Table~\ref{table:downscaled}, when all networks have roughly the same number of parameters, \textit{SegCaps} outperforms all other methods.

\begin{table}[h!]
\centering
\caption{Experimental results on the UHG dataset using downscaled version of \textit{U-Net} and \textit{Tiramisu} to roughly equal the same number of parameters ($1.4$ M) as \textit{SegCaps}. The value of $k$ (number of feature maps per layer reduction factor) for \textit{U-Net} and \textit{P-HNN} is included in parentheses.}
\label{table:downscaled}
\begin{tabular}{@{}c|cc@{}}
\toprule
Method & Dice ($\% \pm$ std) & HD ($mm \pm$ std) \\ 
\toprule
\textit{U-Net} (orig.) & $88.10 \pm 1.84$ & $44.303 \pm 34.148$ \\
\textit{U-Net} ($4.68$) & $87.57 \pm 2.80$ & $62.006 \pm 62.693$ \\
\midrule
\textit{Tiramisu-103} & $87.67 \pm 1.38$ & $61.227 \pm 54.096$ \\
\textit{Tiramisu-56}& $87.68 \pm 0.96$ & $67.913 \pm 36.190$ \\
\midrule
\textit{P-HNN} (orig.) & $88.64 \pm 0.64$ & $43.698 \pm 24.026$ \\
\textit{P-HNN} ($3.2$) & $86.69 \pm 1.39$ & $82.223 \pm 48.989$ \\
\midrule
\textit{SegCaps} & $\bm{88.92 \pm 0.66}$ & $\bm{37.171 \pm 23.223}$ \\
\bottomrule
\end{tabular}
\end{table}

\subsection{Reconstruction Regularization}\label{sec:abl_recon}

The idea of reconstructing the input as a method of regularization was used in CapsNet by \citet{sabour2017dynamic}. The theory behind this technique and the regularization effect it introduces is similar in nature to the problem of ``mode collapse'' in generative adversarial networks (GANs). When training a generative neural network for a specific task through the backpropagation algorithm, the model ``collapses'' to focusing on only the most prevalent modes in the data distribution. A similar phenomenon occurs when you train a discriminative network for a specific task, the model ``collapses'' to only focus on the most discriminative features in the input data and ignores all others. By mapping the capsule vectors back to the input data, this forces the network to pay attention to more relevant features about the input, which might not be as discriminative for the given task, yet still provide some useful information, as evident by the improved results shown in Table~\ref{table:recon}. A similar results can be seen in VEEGAN by \citet{srivastava2017veegan}, where they help solve the issue of mode collapse in GANs through a reconstructor network which reverses the action of the generator by mapping from data to noise. 

\begin{table}[h!]
\centering
\caption{Examining the effect of the proposed extension of the reconstruction regularization to the task of segmentation.}
\label{table:recon}
\begin{tabular}{@{}c|cc@{}}
\toprule
Method & Dice ($\% \pm$ std) & HD ($mm \pm$ std) \\ 
\midrule
No Recon & $88.58 \pm 1.03$ & $42.345 \pm 21.180$ \\
With Recon & $\bm{88.92 \pm 0.66}$ & $\bm{37.171 \pm 23.223}$ \\
\bottomrule
\end{tabular}
\end{table}

\subsection{Dynamic Routing Iterations}\label{sec:abl_routing}
Since the dynamic routing algorithm chosen for this study is an iterative process, we can investigate the optimal number of times to run the routing algorithm per forward pass of the network. In the original work by \citet{sabour2017dynamic}, they found three iterations to provide the optimal results. As seen in Table~\ref{table:routing}, the number of routing iterations does have an effect on the network's performance, and we find the same result in this study of three iterations being optimal over a set of different numbers of iterations studied. \ADD{Several other recent studies have also found three routing iterations to achieve optimal performance~\mbox{\citep{lalonde2020diagnosing,lalonde2020encoding}}. However, other recent studies have found different routing methods or number of routing iterations to be optimal. Likely, as found by~\mbox{\citet{paik2019capsule}}, capsule networks will likely need to find an improved routing mechanism.}

\begin{table}[h!]
\centering
\caption{Examining the effect of different number of routing iterations (abbreviated as \# Iters) per forward pass of \textit{SegCaps}. In $1,3$, one routing iteration is performed when the spatial resolution remains the same and three iterations are performed when the resolution changes.}
\label{table:routing}
\begin{tabular}{@{}c|ccc@{}}
\toprule
\# Iters & Dice ($\% \pm$ std) & HD ($mm \pm$ std) \\ 
\midrule
$1$ & $88.17 \pm 1.23$ & $67.668 \pm 58.556$ \\
$2$ & $88.58 \pm 1.03$ & $42.345 \pm 21.180$ \\
$3$ & $\bm{88.92 \pm 0.66}$ & $\bm{37.171 \pm 23.223}$ \\
$4$ & $87.72 \pm 1.36$ & $110.901 \pm 71.701$ \\
$1,3$ & $88.11 \pm 1.13$ & $72.877 \pm 54.649$ \\
\bottomrule
\end{tabular}
\end{table}

\section{Discussions \& Conclusion}\label{sec:conclusion}

We propose a novel deep learning algorithm, called \textbf{\textit{SegCaps}}, for biomedical image segmentation, and showed its efficacy in a challenging problem of pathological lung segmentation from CT scans \ADD{and thigh muscle and adipose (fat) tissue segmentation from MRI scans, as well as experiments around the affine equivariance properties of a capsule-based segmentation network}. The proposed framework is the first use of the recently introduced capsule network architecture and expands it in several significant ways. First, we modify the original dynamic routing algorithm to act locally when routing children capsules to parent capsules and to share transformation matrices across capsules within the same capsule type. These changes dramatically reduce the memory and parameter burden of the original capsule implementation and allows for operating on large image sizes, whereas previous capsule networks were restricted to very small inputs. To compensate for the loss of global information, we introduce the concept of ``deconvolutional capsules'' and a deep convolutional-deconvolutional capsule architecture for pixel level predictions of object labels. Finally, we extend the masked reconstruction of the target class as a regularization strategy for the segmentation problem. 

Experimentally, \textbf{\textit{SegCaps}} produces improved accuracy for lung segmentation on five datasets from clinical and pre-clinical subjects, in terms of Dice coefficient and Hausdorff distance, when compared with state-of-the-art networks \textit{U-Net} \citep{ronneberger2015u}, \textit{Tiramisu} \citep{jegou2017one}, and \textit{P-HNN} \citep{harrison2017progressive}. \ADD{For muscle and adipose (fat) tissue segmentation, \textit{SegCaps} can perform on par with \textit{U-Net} while only using a small fraction of the parameters, and outperforms the previous state-of-the-art.} More importantly, the proposed \textbf{\textit{SegCaps}} architecture provides strong evidence that the capsule-based framework can more efficiently utilize network parameters, achieving higher predictive performance while using $95.4\%$ fewer parameters than \textit{U-Net}, $90.5\%$ fewer than \textit{P-HNN}, and $85.1\%$ fewer than \textit{Tiramisu}. To the best of our knowledge, this work represents the largest study in pathological lung segmentation, and the only showing results on pre-clinical subjects utilizing state-of-the-art deep learning methods.

The results of these experiments demonstrate the effectiveness of the proposed capsule-based segmentation framework. This study provides helpful insights into future capsule-based works and provides lung-field segmentation analysis on pre-clinical subjects for the first time in the literature.

\ADD{\section*{Acknowledgments}}
\ADD{This study is partially supported by the NIH grant R01-EB020539.}

\bibliographystyle{model2-names.bst}\biboptions{authoryear}
\bibliography{main}

\appendix

\end{document}